\documentclass[a4paper,11pt]{article}
\pdfoutput=1 

\usepackage{jcappub} 
\usepackage{graphicx} 
\usepackage{subcaption}
\usepackage{caption}
\usepackage{multirow}
\usepackage{booktabs}
\usepackage{gensymb}
\usepackage{float}
\usepackage{xspace}
\usepackage{hanging} 
\usepackage{etoolbox}
\usepackage{orcidlink}

\usepackage[demo,
            export]{adjustbox}

\makeatletter
\newcommand{\thickhline}{%
    \noalign{\ifnum0=`}\fi\hrule height 2pt
    \futurelet\reserved@a\@xhline}
\makeatother

\newcommand{\Qmax}{$Q_{\rm max}$\xspace}
\newcommand{\Tmax}{$T_{\rm max}$\xspace}
\newcommand{\Qmin}{$Q_{\rm min}$\xspace}

\title{Isotropy Test with Quasars Using Method of Smoothed Residuals}

\author[a]{A. Antony,\orcidlink{https://orcid.org/0000-0002-6802-459X}}
\emailAdd{akhilantony91@gmail.com}
\author[a,b]{S. A. Appleby,\orcidlink{0000-0001-8227-9516}}
\emailAdd{stephen.appleby@apctp.org}
\author[c]{W.~L.~Matthewson\orcidlink{0000-0001-6957-772X}}
\author[c]{A. Shafieloo\orcidlink{0000-0001-6815-0337}}

\affiliation[a]{Asia Pacific Center for Theoretical Physics, Pohang, 37673, Republic of Korea}
\affiliation[b]{Department of Physics, POSTECH, Pohang, 37673, Republic of Korea}
\affiliation[c]{Korea Astronomy and Space Science Institute, 776, Daedeokdae-ro, Yuseong-gu, Daejeon
34055, Republic of Korea}

\abstract{To assess the significance and scale dependence of anomalous large scale modes in the CatWISE quasar data, we generate smoothed number density fields on the sphere and study their extreme values -- maximum, minimum, maximum antipodal difference. By comparing these summary statistics to those obtained from random isotropic realisations of the data, we determine the statistical significance of large scale modes as a function of smoothing scale. We perform our analysis using five different versions of the data -- the original quasar map, the maps after separately subtracting the ecliptic bias and the CMB dipole, the map obtained after subtracting both, and the map after subtracting the ecliptic bias and anomalous dipole inferred in \cite{Secrest2021}. We find that the ecliptic-corrected, CMB dipole-removed map exhibits large scale modes that are in tension with random realisations of the data (p-values $p \lesssim 10^{-4}$), over a wide range of smoothing scales $\pi/8 \leq \delta \leq \pi/2$. The most prominent feature in the data is an under-density in the southern galactic plane at $(b,\ell) = (-31^\circ,78^\circ)$, which reaches its highest statistical significance when smoothed on scales $\delta = \pi/6$ ($p = 1.2 \times 10^{-6}$). Notably, the minima statistics align with the maximum antipodal difference statistics, whereas the maxima do not. This suggests that the observed dipole-like behavior in the data is primarily driven by the under-density in the southern sky. The ecliptic corrected, anomalous dipole subtracted map reduces the significance of any residual anisotropic features, but an under-density in the south sky persists with p-value $p =0.0018$. }

\begin{document}

\maketitle

\section{Introduction}

Testing the foundational aspects of the standard cosmological model remains a challenging yet essential endeavor in modern cosmology. Within the FLRW description of our Universe, large scale statistical isotropy (SI) and homogeneity are imposed, which allows us to define spatially averaged energy densities that in turn source an average (monopole) expansion rate. On this background, perturbation theory is applied; the ensemble average of density fluctuations are also taken to be rotationally and translationally invariant. These assumptions are difficult to test; inferring the existence of an anisotropic signal requires directionally-sensitive summary statistics \cite{Hajian:2003qq,Dipanshu:2024zex}, and there are many subtleties associated with tests of homogeneity \cite{Buchert:1999er,Buchert:1995fz,Gasperini:2011us,Verweg:2024lps}. Despite these difficulties, extensive efforts have been made to probe these fundamental symmetries. A range of studies, spanning cosmic microwave background analyses, large-scale structure surveys, and other cosmological observations, have sought to identify possible departures from isotropy and homogeneity \cite{Meegan1996, Blake2002, Wright2010, Singal2011, Colin2011, Gibelyou2012, Rubart2013, Nadathur2013, Tiwari2015, Javanmardi2015, Appleby2015, Lin2016, Colin2017, Tiwari2019, Tiwari2019b, pa2019, Secrest2021, Bengaly2024, Kothari2024, Abghari2024}. The results of these investigations continue to refine our understanding of the large-scale structure of the Universe and test the fundamental assumptions underpinning modern cosmology.

{ In addition to the quasar and radio-source dipoles, a number of unexplained large-scale anomalies have been reported, particularly in the cosmic microwave background (CMB). These include the large-angle power asymmetry and alignments of low-$\ell$ multipoles observed in \citep{Planck:2013kqc,Planck:2018nkj} and subsequent analyses \citep{Schwarz:2015cma,Planck:2015igc,Planck:2019evm,Jones:2023ncn}. Another potentially related local anomaly has been considered recently in \cite{Hansen:2025atx}. While the connection between these features and the quasar anisotropies studied here is not established, the growing number of large-scale statistical anomalies highlights the importance of further, independent testing.}

We can expect some degree of violation of isotropy as a function of scale, even within the standard model. In particular, on small scales, isotropy will be violated by the peculiar motion of objects with respect to the cosmic rest frame\footnote{Assuming that such a rest frame exists.}. The precise consistency between the local bulk motion and the standard model is an open question and requires further study with deeper catalogs \cite{Colin2011,Watkins:2023rll,Whitford:2023oww}. However, isotropy on the largest scales can be tested using certain data because it predicts that the last scattering surface as measured by CMB photons, and also any magnitude-limited sample of distant large scale structure tracer particles, should present a spherical distribution with respect to our observer location, after accounting for all relevant peculiar velocities. The presence of a large dipole in the CMB photon distribution is attributed to our peculiar motion with respect to the last scattering surface\footnote{The velocity and direction of this dipole are $v_{\rm pec} = 369.82 \pm 0.11\, {\rm km}\, {\rm s}^{-1}$, $(\ell, b) = (264.02\degree, 48.25\degree)$ in galactic coordinates, assuming that it is purely kinematic \cite{Planck:2018nkj,Planck:2013kqc}.}, and because the matter distribution in the high redshift Universe should be at rest with respect to the CMB, the application of the same velocity correction to the observed matter field should render it statistically isotropic. Using this idea, recent precision tests of the distribution of quasars on the sky have challenged the existence of a cosmological rest frame. Both CatWISE quasars \cite{Marocco2021} and NVSS radio galaxy data \cite{Condon1998} present an excess dipole in number counts \cite{Secrest:2022uvx,Secrest2021}, with an amplitude that is several times larger than the dipole measured from the CMB. Historically both excess dipoles \cite{Singal2011,Rubart2013,Colin2017,Tiwari2019,Singal2019,Siewert2021,Oayda2024} and a dipole consistent with the CMB \cite{Crawford2009,Gibelyou2012, Rubart2014, Tiwari2016, Darling2022, Wagenveld2023,Mittal:2023xub}
 have been found in large scale structure catalogs, but never with the statistical significance of recent work \cite{Secrest2021}. The result was subsequently confirmed in a number of works \cite{Dam2023,Mittal:2023xub,Oayda:2024hnu}, but was also challenged in \cite{Darling2022,Abghari2024}. The latter argued that additional large-scale features are present in the data and that the sky mask, which covers approximately 50\% of the sky, can introduce mode coupling effects that influence the recovered multipoles and their statistical significance. 

Given how central the assumption of isotropy is to building the standard model, testing it using a variety of methodologies is a rational approach. In this work, we apply a different set of statistics to the CatWISE data, with the goal of exploring the statistical significance of the large scale features. Specifically, we generate density fields by smoothing the quasar point distribution on the sky and examine particular one-point and restricted two-point summary statistics, which are the extreme values (maximum, minimum) of the field. One-point statistics provide complementary information compared to higher points, and are less affected by the presence of a mask. Conversely, there is generically less information in the one point statistics of a field compared to two-point measures (like a pair count). By measuring the extreme field values as we vary the smoothing length, we can build an understanding of the typical scales that are anomalous within the data and the extent to which they are anomalous. 

The paper will proceed as follows. In Sections \ref{sec:data}, \ref{sec:stats} we review the data that will be analysed, and the statistics extracted from the quasars, respectively. We present our results in Section \ref{sec:results} and conclude with a discussion in Section \ref{sec:discuss}.

\section{Data}
\label{sec:data}

For the purpose of our study, we ultilize an all-sky flux-limited quasar sample generated from the Wide-field Infrared Survey Explorer (WISE) mission \cite{Wright2010}. WISE surveyed the sky in four infrared bands $W1 (3.4\mu {\rm m})$, $W2 (4.6\mu {\rm m})$, $W3 (12 \mu {\rm m})$, $W4 (22\mu {\rm m})$, and sources were selected using $W1$ and $W2$ due to their superior depth. We use the catalog constructed by Secrest et al. \cite{Secrest2021}. The data set comprises 1,355,352 quasars selected from the CatWISE2020 catalog. The selection focuses primarily on emission at wavelengths of $3.4 \mu m$ (W1) and $4.6 \mu m$ (W2) with a color cut of $W1 - W2 > 0.8$ applied to ensure that AGN-dominated emission follows a power-law distribution. By cross-correlating the quasar sample with the eBOSS sample \cite{eBOSS:2015jyv}, it was found that the CatWISE data has a mean redshift of $z \sim 1.2$ \cite{Secrest2021}, making it largely insensitive to local bulk velocity effects. 

\begin{figure}
\includegraphics[width=0.32\linewidth]{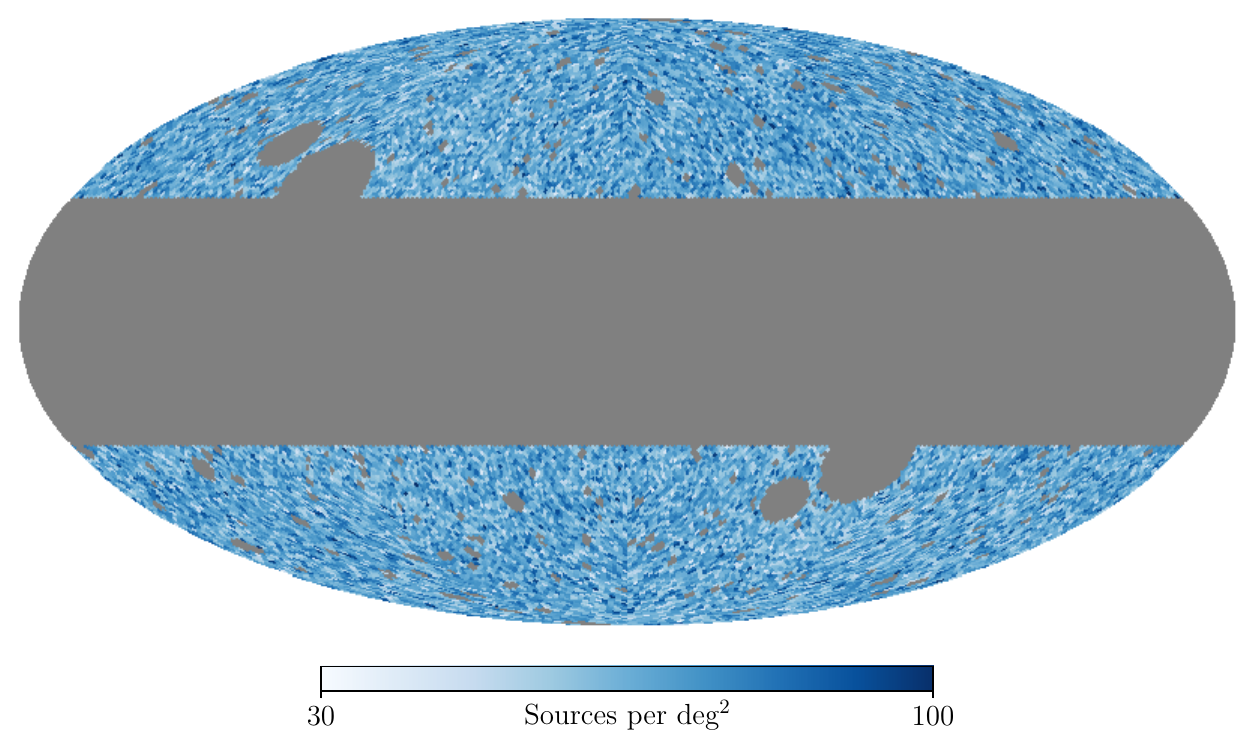}  
\includegraphics[width=0.32\textwidth]{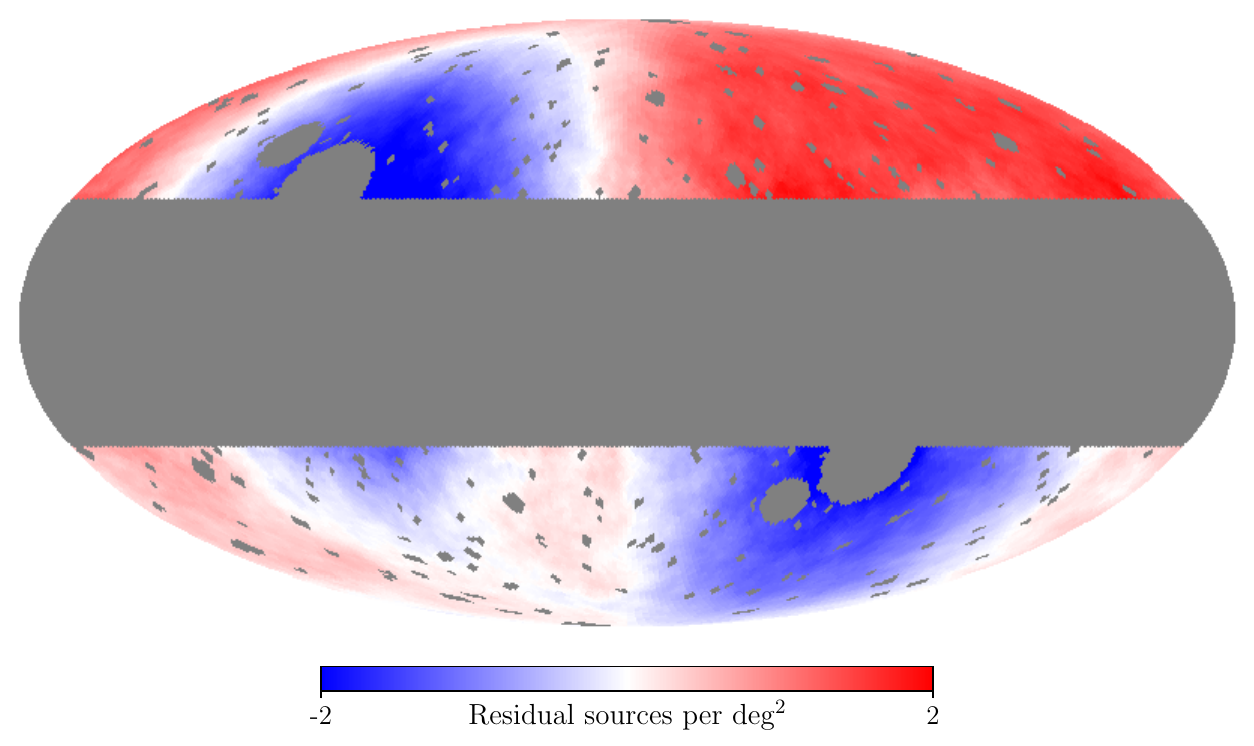}
\includegraphics[width=0.32\textwidth]{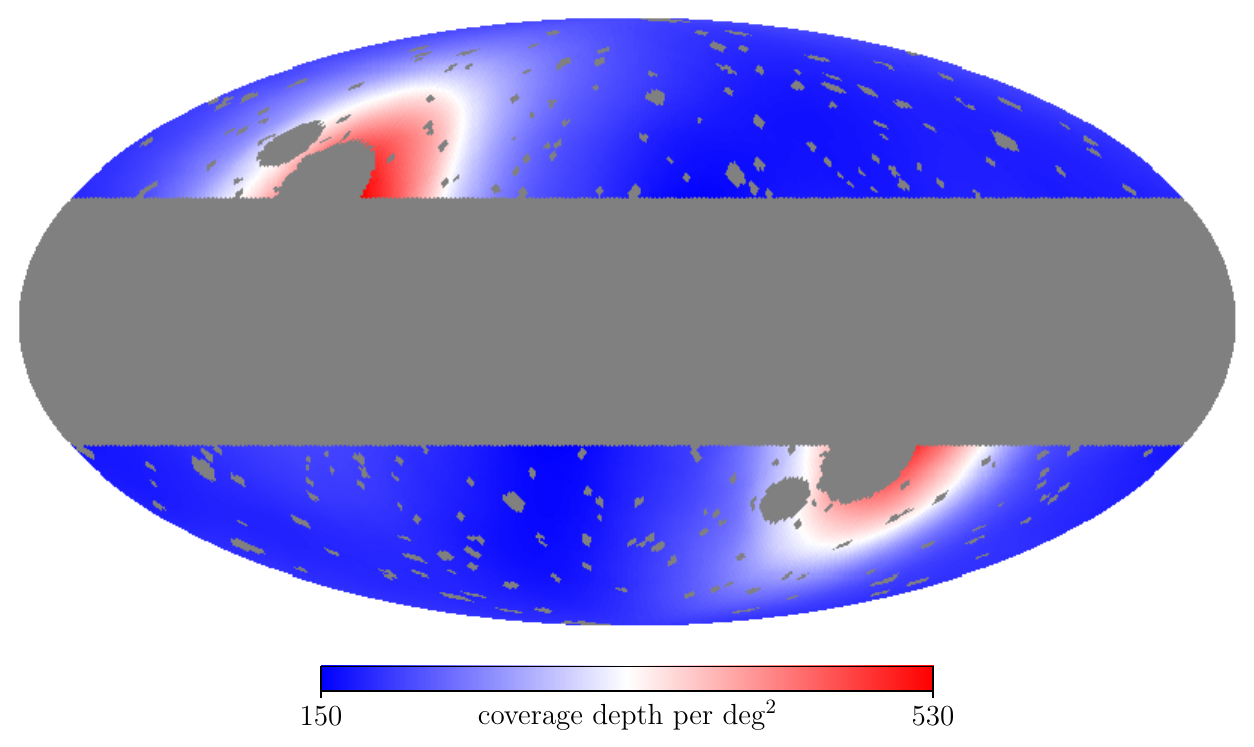}
\caption{[Left panel] CatWISE quasar data binned into a $N_{\rm side} = 64$ HEALPix map, unsmoothed and mean subtracted. [Middle panel] The same data as in the left panel, but now smoothed using a tophat filter of area 1 steradian. [Right panel]
The $W1$ coverage map, also smoothed with a 1 steradian tophat.  }
\label{fig:data}
\end{figure}

The quasars are binned into a HEALPix\footnote{ http://healpix.sourceforge.net} \cite{Gorski:2004by} map with $N_{\rm side} = 64$; the mean subtracted data is presented in the left panel of Figure \ref{fig:data}. The same data, tophat smoothed with smoothing area 1 steradian, is presented in the middle panel and the $W1$ coverage depth of the survey, also smoothed over 1 steradian, is presented in the right panel. The quasar distribution presents a clear correlation with the scanning of the telescope. We discuss the removal of this effect in Section \ref{sec:eclcor_res}.

The grey regions in Figure \ref{fig:data} are masked. We use the same mask adopted in (Secrest et. al), which cuts all sources below $|b| <30\degree$ and removed areas of poor photometry or artifacts using masks of radius $< 2\degree$. Additionally, it also masks the Magellanic Clouds and Andromeda. In total, 291 regions in the sky, excluding the galactic plane, have been removed from the analysis. In an attempt to minimize any potential bias, we apply both the mask and antipodal image (in galactic coordinates) to the data. This is because one of the statistics that we measure is defined as the difference in density between antipodal points, and we want to avoid ambiguous situations where one pixel is masked\footnote{The additional masking does not significantly affect our conclusions. This is expected because the original mask is almost entirely antipodally symmetric (the galactic plane) and the other features; most notably LMC and SMC, are small in angular extent relative to the smoothing scales adopted in this work}. In total, $54\%$ of the sky is masked in our analysis. The data is corrected further by subtracting the CMB dipole and also a quadrupole that is assumed to be due to the scanning strategy of the telescope. These steps are documented in Section \ref{sec:results}. 

If the sources that we are measuring exhibit a power law spectral energy distribution $S \propto \nu^{\alpha}$, and the cumulative number count as a function of limiting flux density has the form $N(>S) \propto S^{-x}$, then our peculiar motion (magnitude $v$) with respect to the assumed isotropic sample will generate a dipole in the observed number count of magnitude

\begin{equation}\label{eq:dip} \mathcal{D} = \left[ 2 + x (1+\alpha) \right]{v \over c} + {\cal O}\left({v^{2} \over c^{2}}\right) . \end{equation} 

\noindent Any flux limited sample with the above properties can therefore be used to test statistical isotropy, and specifically the standard model posits that any distant collection of objects should be at rest with respect to CMB photons, such that $v \simeq 369.82 {\rm km} \, {\rm s}^{-1}$ in equation (\ref{eq:dip}). This was first discussed by Ellis and Baldwin \cite{Ellis1984} and  is a generic prediction for isotropic cosmological models.

\section{Methodology}
\label{sec:stats}

In this Section we review the summary statistics that we extract from the quasar data, and how they can be used to determine the nature and significance of large scale features in the catalog.

To quantify the level of anisotropy in the quasars, we create a smoothed residual map of the data. Statistical measures are computed from these residuals and their significance is evaluated against mock datasets that assume isotropy. This study focuses on two specific summary statistics: the $Q$- and the $T$-measures. To compute these quantities, we start by defining a field $Q(\theta,\phi)$, which is generated by smoothing the quasar number density. The procedure for this computation is outlined below:

\begin{enumerate}
    \item \textbf{Construct a density map}: Create a density map by binning the quasars into a Healpix map. The monopole is subtracted from the map to obtain the residuals, which are represented as:
    \begin{equation}
        q(\hat{n}) = q_0(\hat{n}) - \bar{q},
    \end{equation}
    where $q_0(\hat{n})$ is the original density, and $\bar{q}$ is the mean density value. $\hat{n}$ is the unit-normal pointing to the pixels. 

    \item \textbf{Generate a mask}: Generate a binary mask, $m(\hat{n})$, at the same resolution as the density map.

    \item \textbf{Compute the smoothed data values}: For each unmasked pixel in the density map, calculate the smoothed value, $Q_{\text{data}}$, using a specified smoothing width, $\delta$, as follows:
    \begin{align}
        Q_{\text{data}}(\theta,\phi;\delta) &= \sum_{i=1}^{N} q_i(\theta_i,\phi_i) W(\theta,\phi,\theta_i,\phi_i;\delta), \\
        W(\theta,\phi,\theta_i,\phi_i;\delta) &= \frac{1}{\sqrt{2\pi}\delta} \exp\bigg[-\frac{L(\theta,\phi,\theta_i,\phi_i)^2}{2\delta^2}\bigg]\Theta(\delta-L),
    \end{align}
    where the angular distance $L(\theta,\phi,\theta_i,\phi_i)$ between pixel at angular position $(\theta,\phi)$ and the $i^{\rm th}$ pixel at $(\theta_{i},\phi_{i})$ is given by:
    \begin{align}
        L(\theta,\phi,\theta_i,\phi_i) &= 2 \arcsin \frac{R}{2}, \\
        R &= \Big(\big[\sin(\theta_i)\cos(\phi_i) - \sin(\theta)\cos(\phi)\big]^2 \\
        &+ \big[\sin(\theta_i)\sin(\phi_i) - \sin(\theta)\sin(\phi)\big]^2 \\
        &+ \big[\cos(\theta_i) - \cos(\theta)\big]^2\Big)^{1/2} ,
    \end{align}
  and the step function is defined as $\Theta(\delta - L) = 1$ if $L < \delta$ and zero otherwise.

    \item \textbf{Compute the smoothed mask values}: Similarly, calculate the smoothed mask value, $Q_{\text{mask}}$, for each pixel using the same smoothing width:
    \begin{equation}
        Q_{\text{mask}}(\theta,\phi;\delta) = \sum_{i=1}^{N} m_i(\theta_i,\phi_i) W(\theta,\phi,\theta_i,\phi_i;\delta ).
    \end{equation}

    \item \textbf{Calculate the scaled $Q$-value}: Normalize the smoothed data values by the smoothed mask values to compute the scaled $Q$-value:
    \begin{equation} \label{qscaled}
        Q_{\text{scaled}} = \frac{Q_{\text{data}}}{Q_{\text{mask}}}.
    \end{equation}

    \item \textbf{Calculate the $T$-value}: The field $T(\theta,\phi)$ is defined as the difference between the $Q$-value at a given pixel and the $Q$-value at its antipodal pixel
    \begin{equation}
    T(\hat{n}) =  Q_{scaled}(\hat{n}) - Q_{scaled}(-\hat{n})\,.
    \end{equation}
\end{enumerate}

The method is straightforward and in this context describes the generation of a smoothed density field from a pixelated map in real space. It has been more generally used to search for anisotropic signals in sparse data \cite{Colin2011,Appleby2015}. The particular smoothing kernel adopted in this work is a product of a top hat and Gaussian. We selected Gaussian smoothing because it weights pixels by separation, but when selecting large $\delta \sim \pi/2$ scales the Gaussian kernel can over-smooth the field, potentially washing out the largest scale feature in the data (a dipole). This motivates the implementation of a top hat cut at $L = \delta$. 

From the $Q$ map constructed above, we find the maximum $Q_{\rm max}$ and minimum $Q_{\rm min}$ values and also the maximum $T_{\rm max}$ of the $T$ map; it is these statistics that will be used to quantify the magnitude and significance of features in the data as a function of scale $\delta$. Evaluating the significance of extreme values of a density field is a well established methodology in cosmology \cite{Colin2011,Appleby2015}. In $Q_{\rm min}$ and $Q_{\rm max}$, we are studying the one-point statistics of the field. $T_{\rm max}$ is a two-point statistic, but is sensitive only to odd, large-scale multipoles. The relative location of $Q_{\rm max}$, $Q_{\rm min}$ and $T_{\rm max}$ on the sphere provides additional information about the nature of the density field. If a dipole dominates the large scale modes, one would expect $Q_{\rm max}$ and $Q_{\rm min}$ to be antipodal and aligned with the \Tmax direction. 

To determine the statistical significance of $Q_{\rm max}$, $Q_{\rm min}$ and $T_{\rm max}$, we need to calculate these same statistics for a set of isotropic mock data. To do so we generate $N_{\rm real}=10^{5}$ mock realisations, in which the original $q(\hat{n})$ quasar map is randomly scattered into the unmasked pixels and $Q_{scaled}$ and $T$ maps are constructed using the smoothing algorithm above. $Q_{\rm max}$, $Q_{\rm min}$ and $T_{\rm max}$ are extracted from each random, isotropic realisation and a probability distribution is generated. The statistical significance of the measured $Q_{\rm max}$, $Q_{\rm min}$, $T_{max}$ values are inferred from these mock probability distributions. The procedure is repeated for various smoothing scales $\pi/2 \leq \delta \leq \pi/16$. { However, for the case $\delta = \pi/6$, we generated $3\times10^{6}$ mock realisations to compute the exact significance, as the signal was found to be strongest at this smoothing scale.}

From the $Q$ maps, we also extract the two-point angular correlation function to highlight the dominant large scale features in the data. The Landy-Szalay(LS) estimator \cite{Landy1993} is a widely used and robust method for calculating the angular two-point correlation function, and is given by
\begin{equation}
    \zeta(\theta) = \frac{DD(\theta)-2DR(\theta)+RR(\theta)}{RR(\theta)}\,,
\end{equation}

\noindent where $DD$ is the number of data-data pairs at angular separation $\theta$, $DR$ is the number of data-random pairs and $RR$ is the number of random-random pairs. The random catalogs are generated by scattering the residual pixels ($q$) on the sphere while preserving the mask. The estimator minimizes edge effects and biases present in simpler estimators, and also has lower variance compared to other estimators.

\begin{figure}
\begin{minipage}{.5\linewidth}
\centering
\includegraphics[width=.98\linewidth,valign=c]{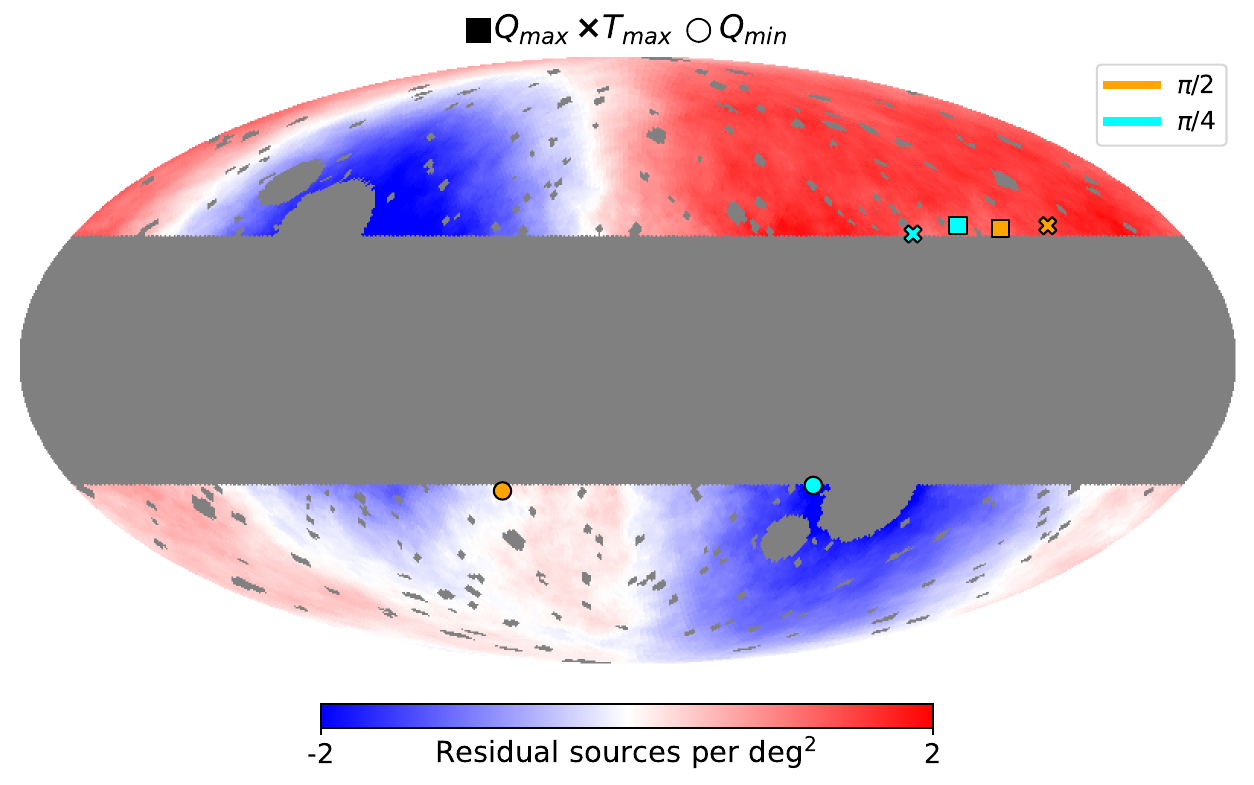} 
\end{minipage}
\begin{minipage}{.5\linewidth}
\centering
\includegraphics[width=.98\linewidth]{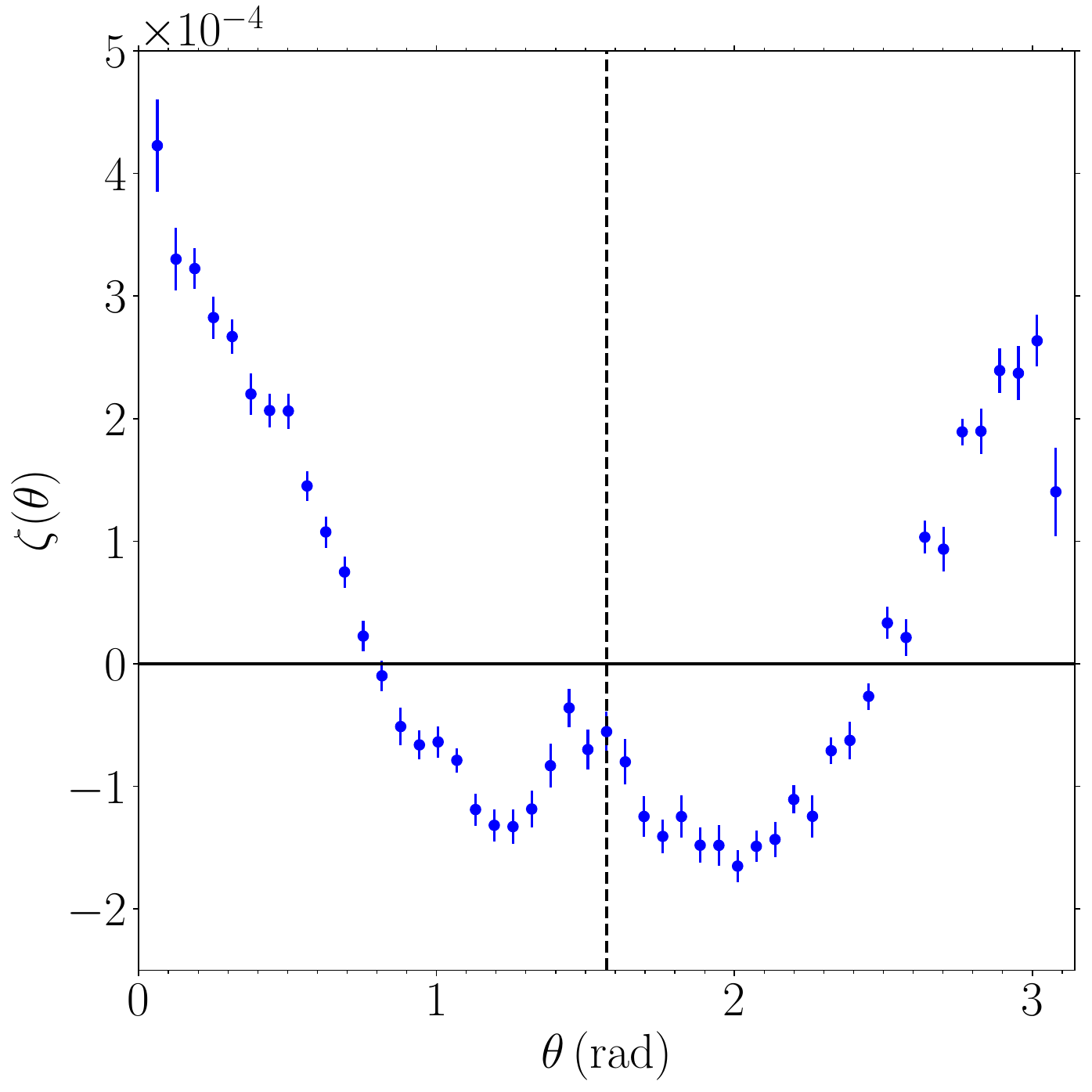}
\end{minipage} 
\\
\includegraphics[width=\textwidth]{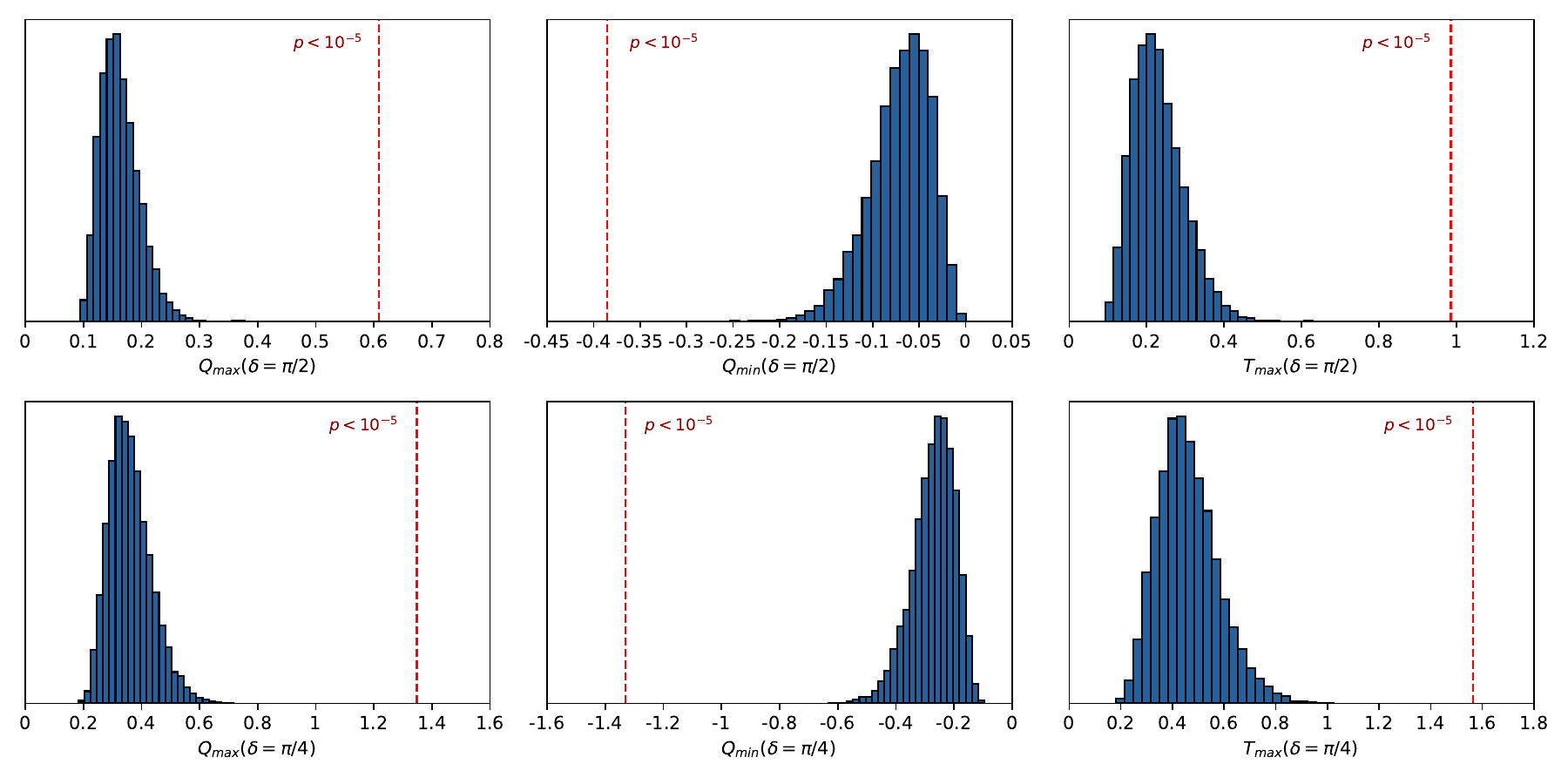}
\caption{
Results from the real-space analysis of the full data density map. 
\textbf{Top Left:} The smoothed density map of the quasar distribution from the CatWISE catalog. The gray regions represent masked areas in the map. 
\textbf{Top Right:} The angular correlation function, $\zeta(\theta)$, where $\theta$ is in radians. The shape is predominantly quadrupolar but is modulated by a dipole. 
\textbf{Bottom Rows:} The distributions of \Qmax, \Qmin and \Tmax from $10^{5}$ isotropic realisations are shown (blue histograms) alongside the corresponding values of the same statistics measured from the data (vertical red dashed lines). Results are presented for two smoothing scales: $\delta = \pi/2$ (middle row) and $\delta = \pi/4$ (bottom row).
}
\label{fig:full_data}
\end{figure}

\section{Results}
\label{sec:results}

In this section, we present the findings of our analysis. Using the methodology outlined in Section \ref{sec:stats}, we systematically examine the quasar datasets by sequentially removing known signals from the complete dataset and analyzing the resulting density maps.

\subsection{Full Data Density Map}

The analysis of the complete, unmodified CatWISE dataset is presented in \autoref{fig:full_data}. This dataset predominantly exhibits two signals: a dipole and a quadrupole. In the top left panel we present the smoothed density field on the sky, in the top right panel the angular correlation function, and in the lower panels the statistics $Q_{\rm max}, Q_{\rm min}$ and $T_{\rm max}$ (left, middle, right columns) for two smoothing scales; $\delta = \pi/2$ (top row) and $\delta = \pi/4$ (bottom row). The vertical red dashed lines are the values of the statistics extracted from the data, and the blue histograms are $N=10^{5}$ isotropic realisations that we use to determine the statistical significance. The angular correlation function clearly identifies a strong quadrupole signal, that is modulated by a dipole. The latter manifests as an amplitude difference between the maxima of the angular correlation function. There is also a feature in the form of a bump at angular separations $\theta \sim \pi/2$. 

{ We find a dipole signal that is highly significant : $p < 10^{-5}$ for \Qmax,  \Qmin  and \Tmax at a smoothing width of $\delta = \pi/2$, based on $10^{5}$ isotropic realisations).} This dipole signal is typically associated with the relative motion of the observer with respect to the CMB rest frame. { At a reduced smoothing width of $\delta = \pi/4$ (bottom panels), we observe that the data values of \Qmax and \Qmin are also anomalous at the level $p < 10^{-5}$ compared to the isotropic histograms.} For both $\delta = \pi/2$ and $\pi/4$, we find zero isotropic realisations that generate \Qmax, \Qmin or \Tmax that are comparable to the data. This first case simply serves as a validation of our methodology; visual inspection of the map shows that it is dominated by a quadrupole {\bf that is modulated by a dipole. }

{ We note that because the density fields are smoothed after mean subtraction, $Q_{\rm min}$ and $Q_{\rm max}$ are not symmetric about zero in the isotropic realisations. This asymmetry arises from the presence of the mask, which modifies the distribution of pixel values after smoothing. This has no impact on the analysis, since the same procedure is applied consistently to both the data and the isotropic realisations.}

\begin{figure}
\begin{minipage}{.5\linewidth}
\centering
\includegraphics[width=.98\linewidth,valign=c]{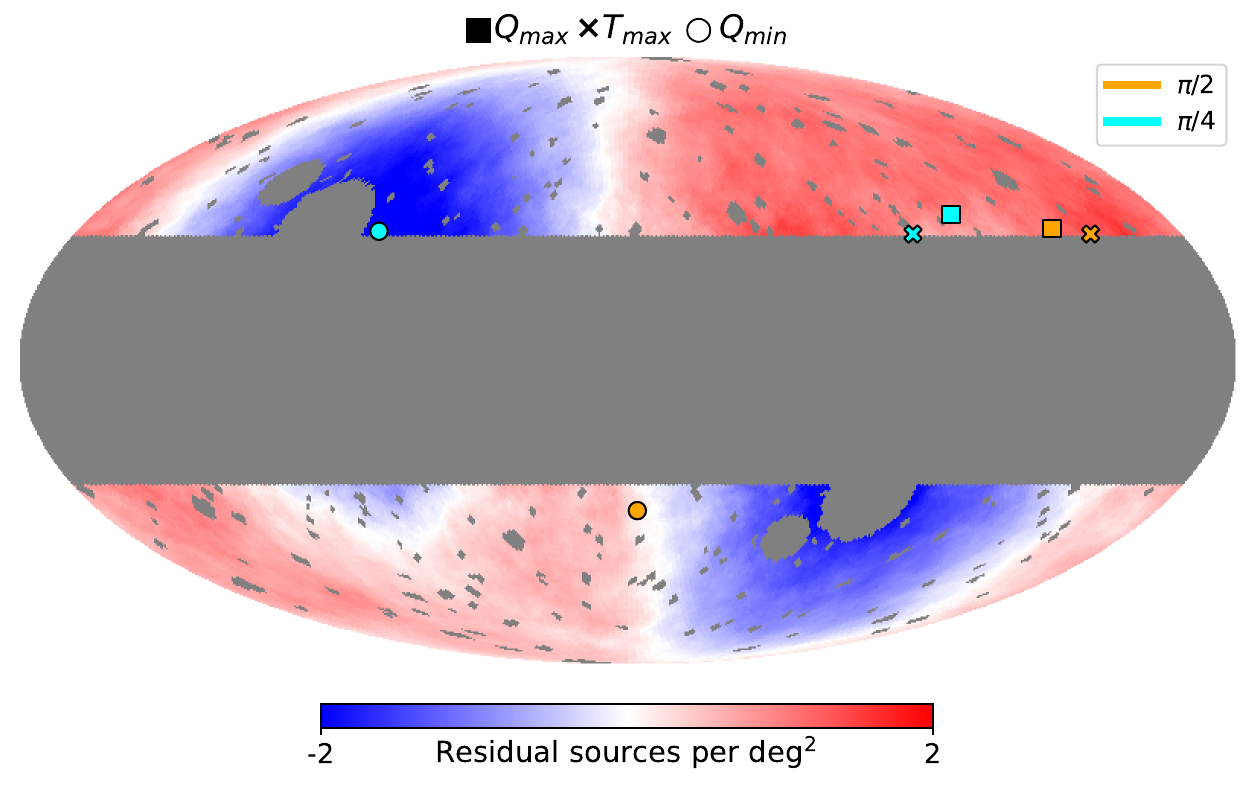} 
\end{minipage}
\begin{minipage}{.5\linewidth}
\centering
        \includegraphics[width=\textwidth]{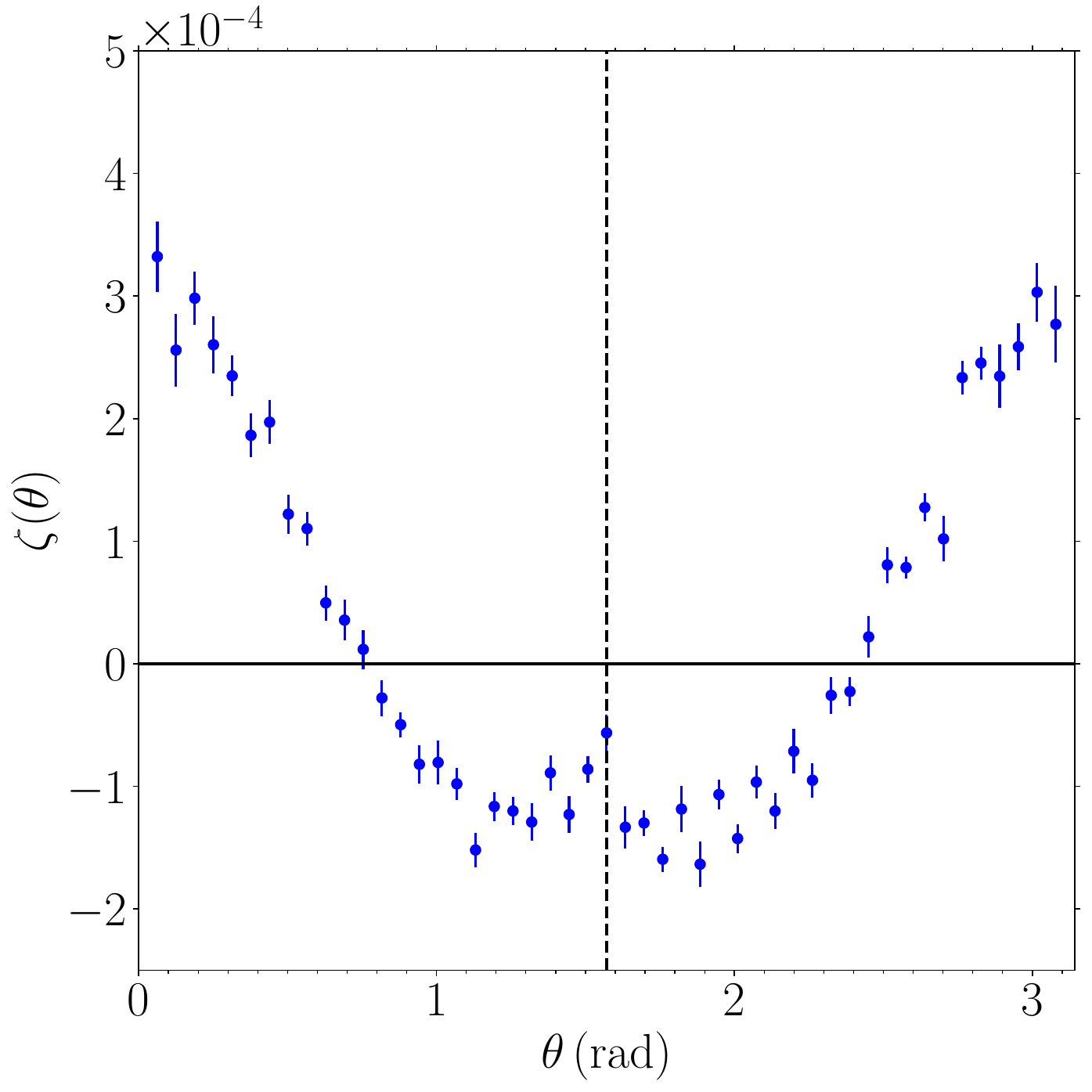}
\end{minipage} 
\\
        \includegraphics[width=\textwidth]{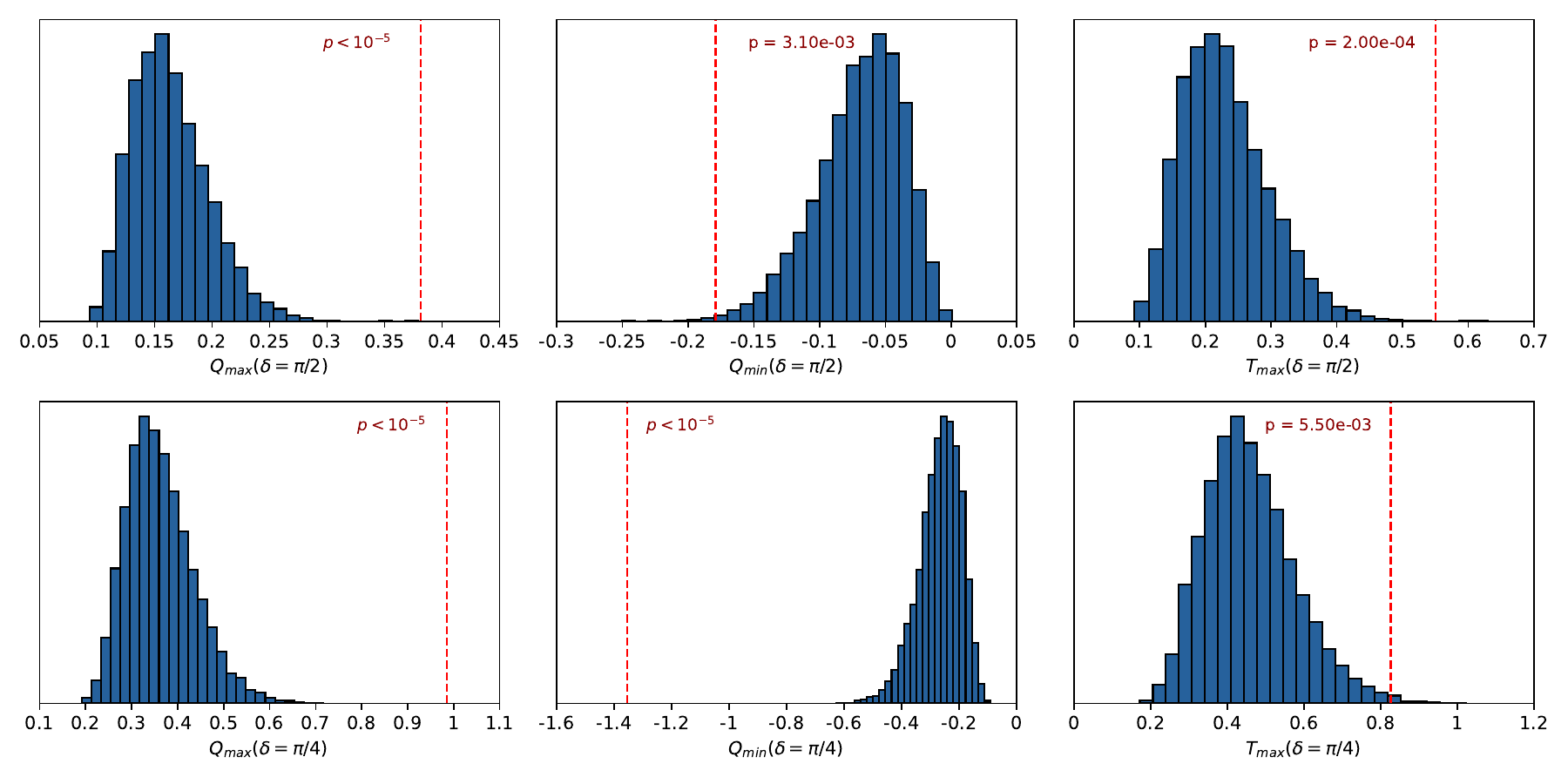}
\caption{
Results from the real-space analysis of the CMB dipole-removed density map. 
\textbf{Top Left:} The smoothed density map of the quasar distribution from the CatWISE catalog. The gray regions represent masked areas in the map. 
\textbf{Top Right:} The angular correlation function, $\zeta(\theta)$, where $\theta$ is in radians. The shape is now predominantly quadrupolar. 
\textbf{Bottom Rows:} The distributions of \Qmax, \Qmin and \Tmax from $10^{5}$ isotropic realisations are shown (blue histograms) alongside the corresponding values of the same statistics measured from the data {(vertical red dashed lines)}. Results are presented for two smoothing scales: $\delta = \pi/2$ (middle row) and $\delta = \pi/4$ (bottom row).
}
\label{fig:dm_nocmb_ps}
\end{figure}

\subsection{Full Data Density Map After Removing the CMB Dipole Signal}

The quasar map contains at least one large-scale feature that is expected within the standard model; our velocity with respect to the CMB, and presumably large scale structure, should generate a dipole in the number count \cite{Ellis1984}. To further isolate signals, we subtract the CMB dipole contribution from the full dataset and repeat the analysis. The method by which we subtract the dipole is described in \autoref{dipole_removal}. The results are presented in \autoref{fig:dm_nocmb_ps}; the panels show the same quantities as in \autoref{fig:full_data}. The angular correlation function reveals a strong quadrupole signal alongside a much reduced, but still non-zero, dipole modulation. The quadrupole is clearly still the dominant signal in the map. 

In the lower panels, we see that upon removing the CMB dipole signal the significance of \Qmax, \Qmin  and \Tmax  decreases at $\delta=\pi/2$. At $\delta=\pi/4$, \Qmax  and \Qmin remain highly significant $p < 10^{-5}$, while \Tmax  shows a reduced significance of $p = 0.011$. This difference arises because on large scales \Tmax is sensitive solely to a dipole signal, and the dominant dipole component, the CMB dipole, has been removed. The remaining significant signal may be attributed to an excess dipole intrinsic to the quasar dataset. \Qmin and \Qmax are sensitive to the quadrupole and remain highly significant, with zero isotropic realisations being found with comparable values to the data. 

\subsection{Ecliptic-Corrected Density Map}
\label{sec:eclcor_res}

\begin{figure}[t]
\begin{minipage}{.5\linewidth}
\centering
\includegraphics[width=.98\linewidth,valign=c]{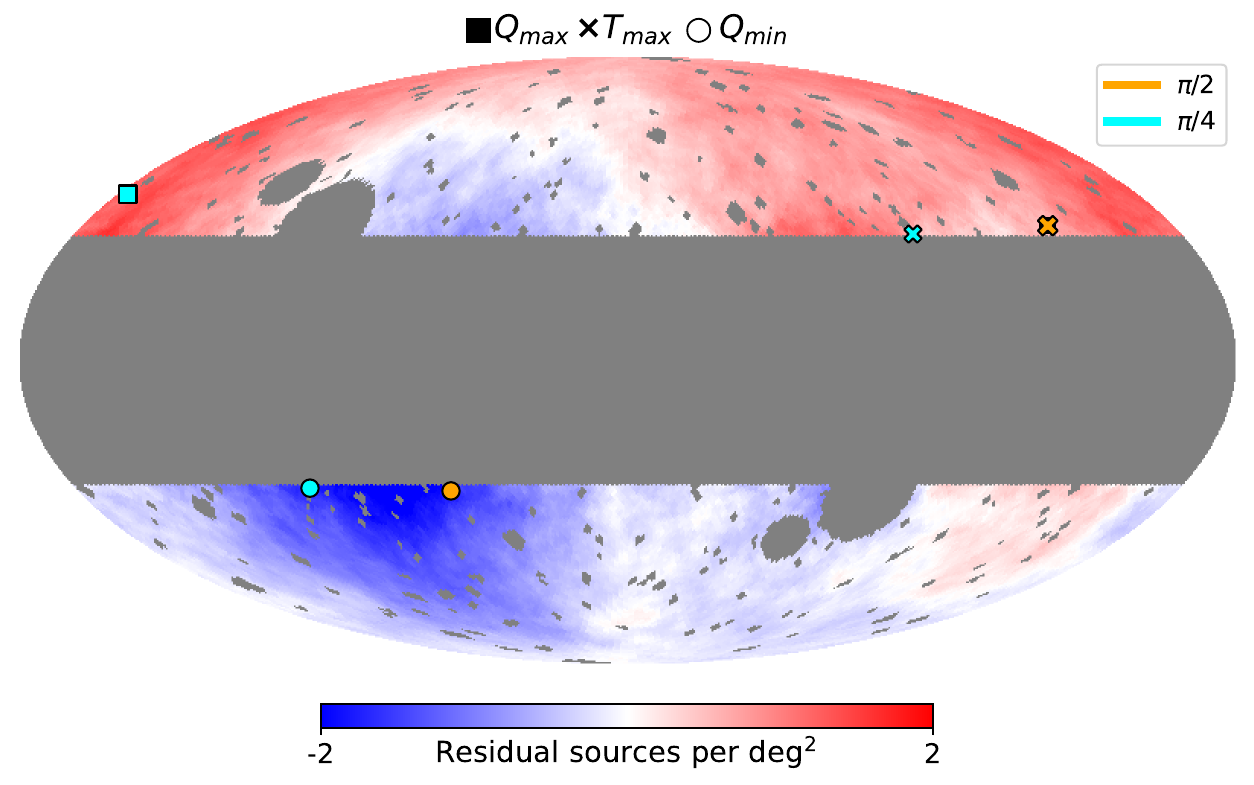}
\end{minipage}
\begin{minipage}{.5\linewidth}
\centering
        \includegraphics[width=\textwidth]{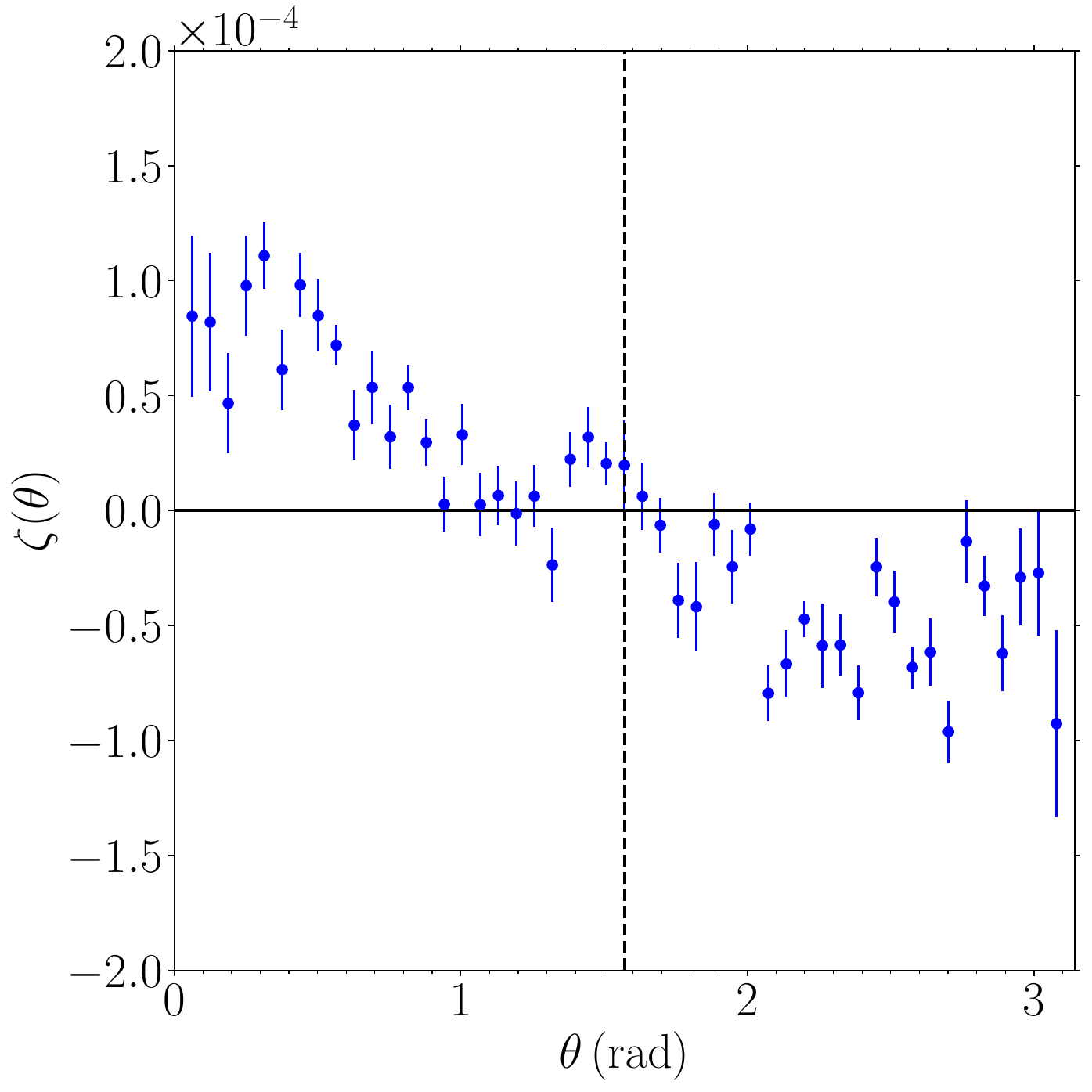}
\end{minipage} 
\\
        \includegraphics[width=\textwidth]{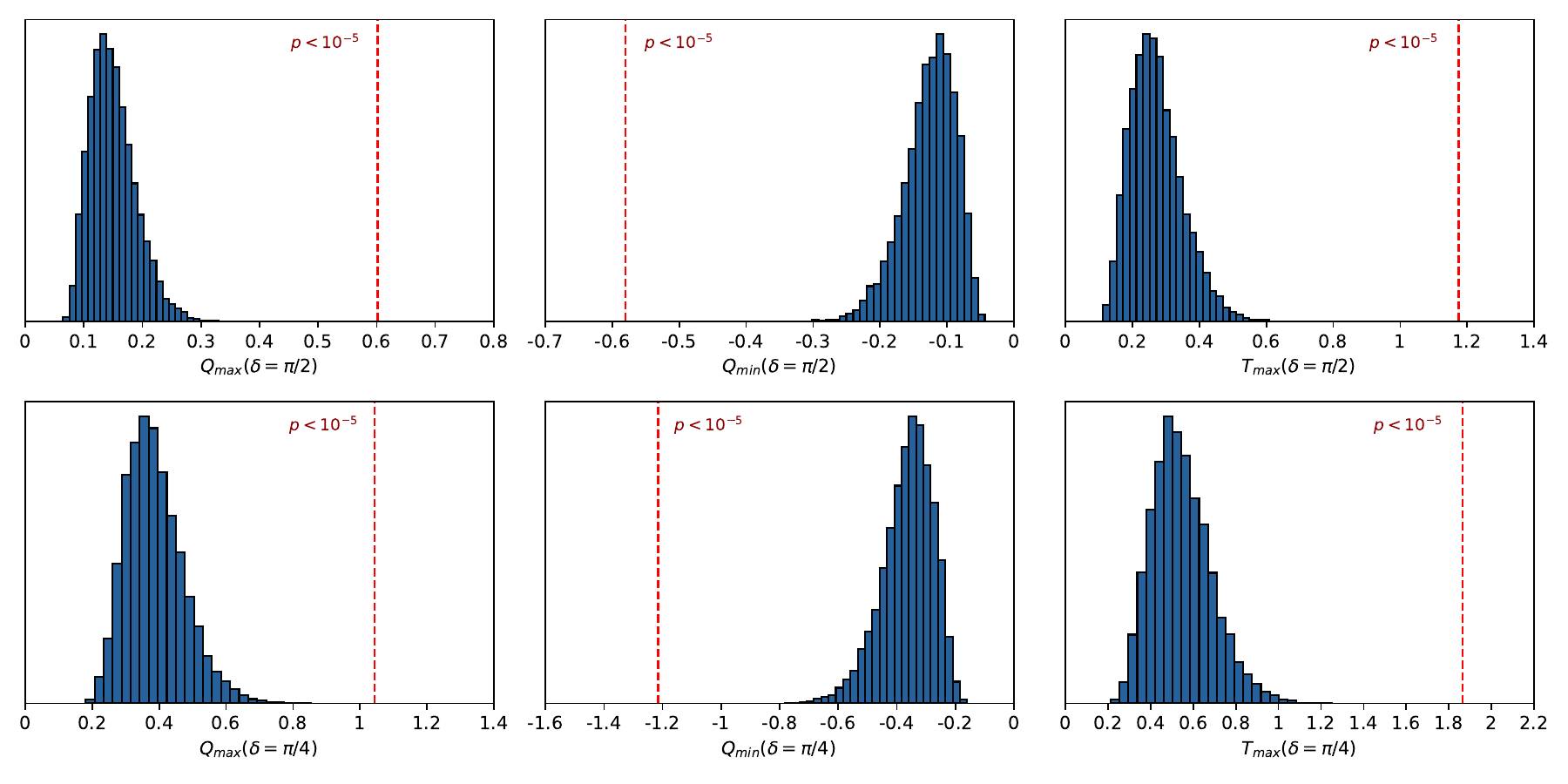}
\caption{\label{fig:eclcor_data}Results from the real-space analysis of the ecliptic-corrected density map. 
\textbf{Top Left:} The smoothed density map of the quasar distribution from the CatWISE catalog. The gray regions represent masked areas in the map. 
\textbf{Top Right:} The angular correlation function, $\zeta(\theta)$, where $\theta$ is in radians. The shape is predominantly dipolar.
\textbf{Bottom Rows:} The distributions of \Qmax, \Qmin and \Tmax from $10^{5}$ isotropic realisations are shown (blue histograms) alongside the corresponding values of the same statistics measured from the data {(vertical red dashed lines)}. Results are presented for two smoothing scales: $\delta = \pi/2$ (middle row) and $\delta = \pi/4$ (bottom row).}
\end{figure}

The presence of a quadrupole is well known, and is typically attributed to the scanning strategy of the telescope. We correct for the ecliptic scanning pattern of WISE as described in \cite{Secrest2021,Secrest:2022uvx}; the quasar number count varies approximately linearly with ecliptic latitude. The data is binned in ecliptic latitude, a linear curve (slope= $-0.051$ , intercept= 68.89 $\deg^{-2}$ ) is fitted to the number counts, and sources are artificially added to the map to remove the linear systematic variation. The result of this procedure is presented in \autoref{fig:eclcor_data}. The density map (top left panel) still presents large scale modes, most clearly a prominent dipole signal. This is partly due to the CMB dipole, which is also identified in the correlation function (top right panel). 

The \Qmin, \Qmax and \Tmax statistics are now consistent with the dominance of a dipole signal. At $\delta = \pi/2$ smoothing,  \Qmax, \Qmin and \Tmax each exhibit high significance. At $\delta = \pi/4$, the significance of the data relative to the mocks shows no notable change. If a $\sim \pi/4$ feature was present in the map, then we would observe a change in the significance of \Qmax and \Qmin relative to the $\pi/2$ smoothing, but not in \Tmax. The fact that we observe no significant change in the significance of any of the statistics indicates that the quadrupole signal has been successfully removed by the ecliptic correction. The exact reason why the data should be cleaned in this way, and why there is a linear relation between the number counts as a function of ecliptic latitude, is uncertain. Stellar contamination has been proposed as an explanation \cite{Abghari2024}, or alternatively bright obfuscating faint sources in regions where the highest number of observations are made.

\begin{figure}[htbp]
\begin{minipage}{.5\linewidth}
\centering
\includegraphics[width=.98\linewidth,valign=c]{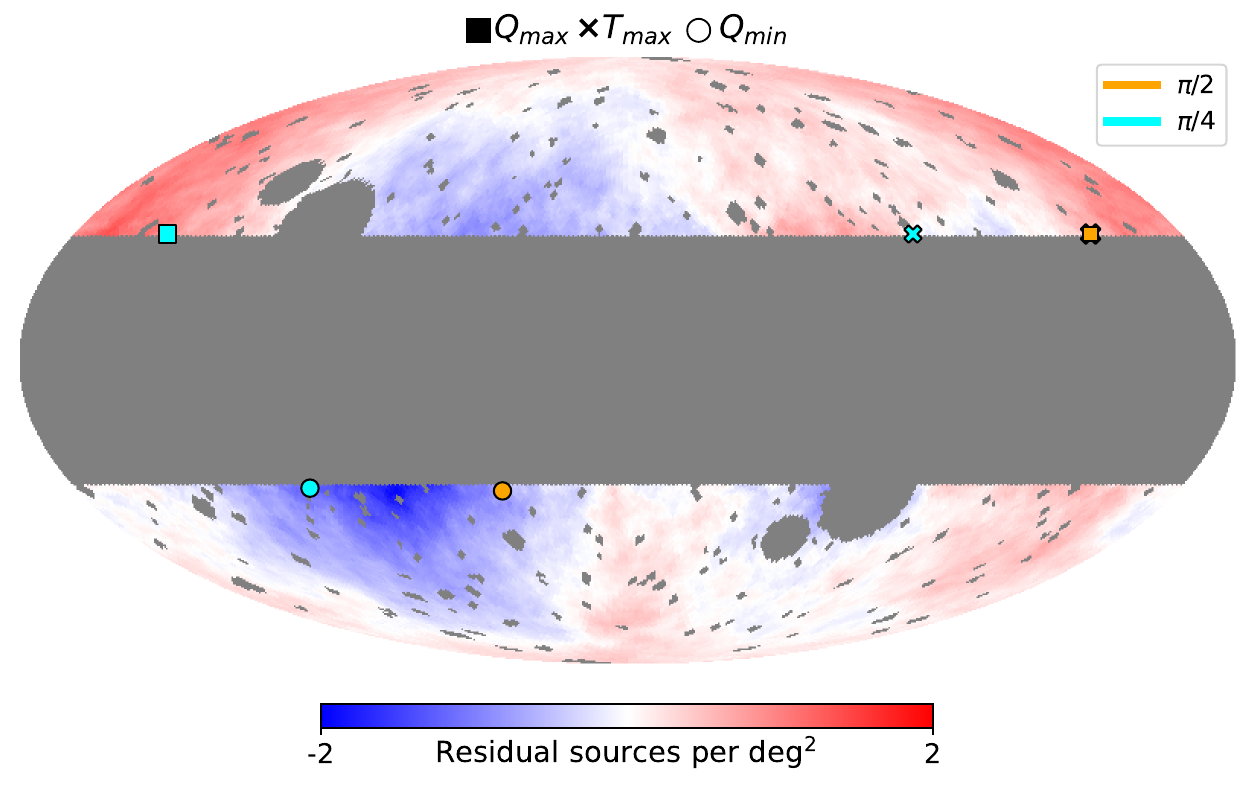}
\end{minipage}
\begin{minipage}{.5\linewidth}
\centering
        \includegraphics[width=\textwidth]{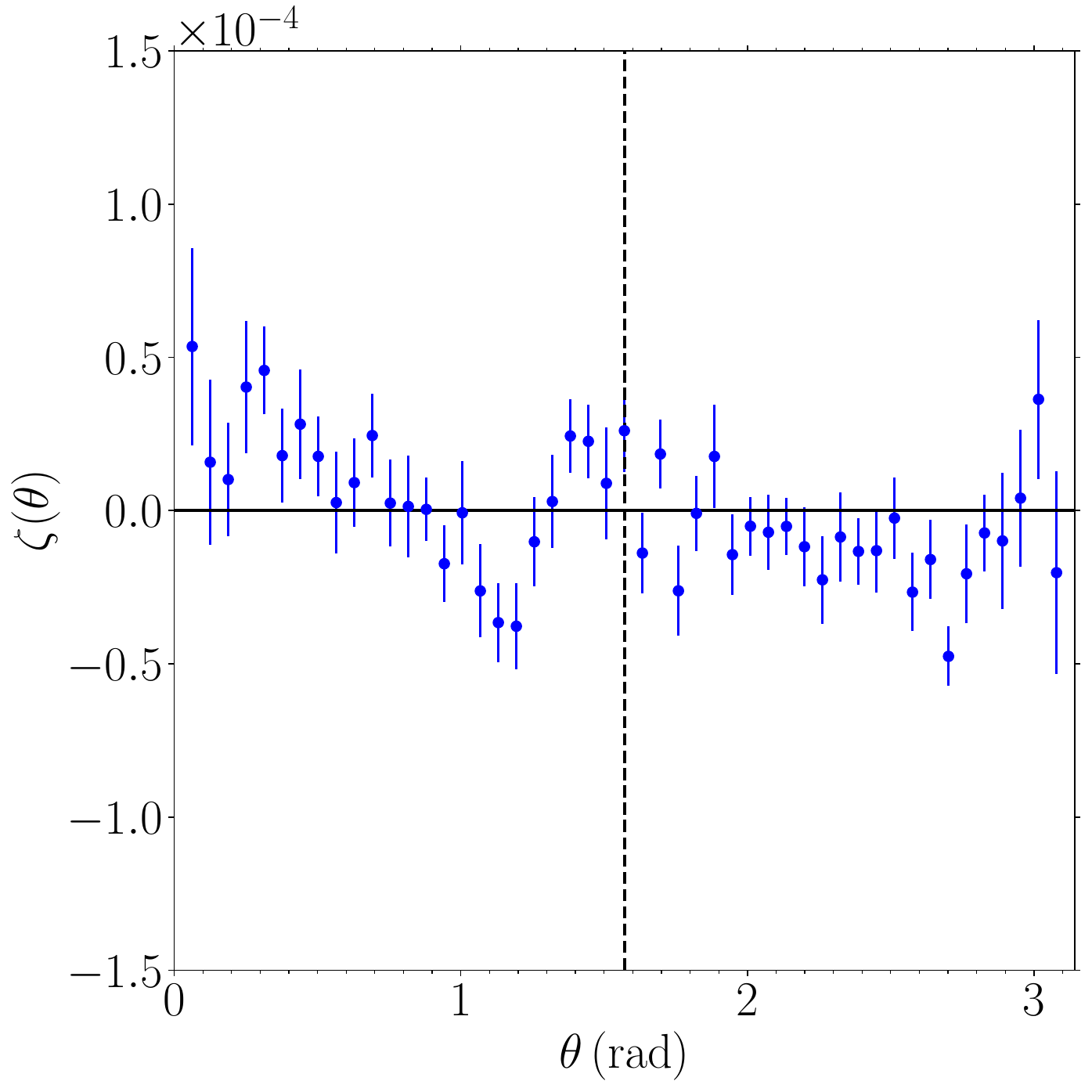}
\end{minipage} 
\\
        \includegraphics[width=\textwidth]{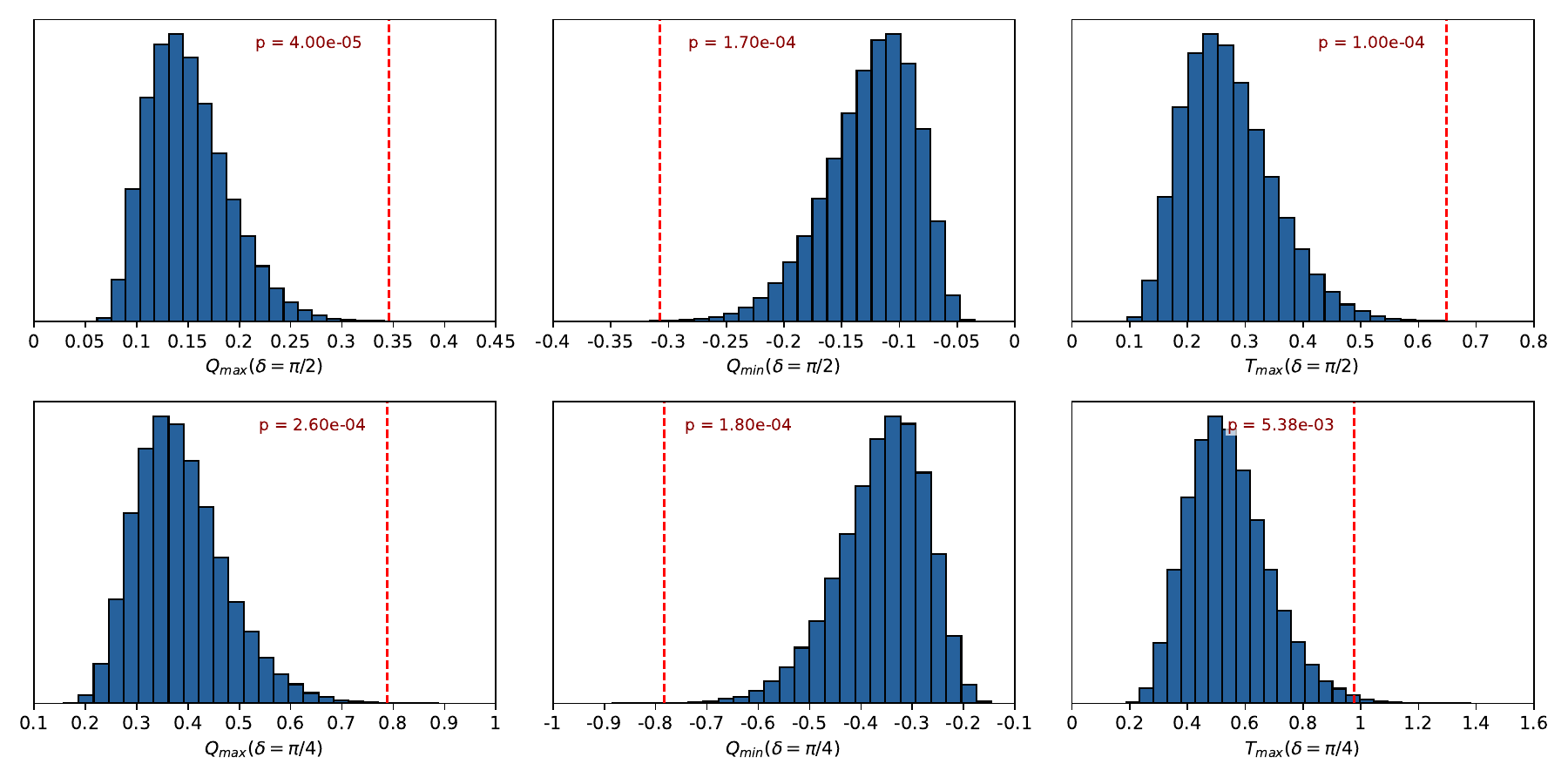}
\caption{\label{fig:eclcor_nocmb_data}Results from the real-space analysis of the CMB dipole-removed and ecliptic-corrected density map. 
\textbf{Top Left:} The smoothed density map of the quasar distribution from the CatWISE catalog. The gray regions represent masked areas in the map. 
\textbf{Top Right:} The angular correlation function, $\zeta(\theta)$, where $\theta$ is in radians. 
\textbf{Bottom Rows:} The distributions of \Qmax, \Qmin and \Tmax from $10^{5}$ isotropic realisations are shown (blue histograms) alongside the corresponding values of the same statistics measured from the data {(vertical red dashed lines)}. Results are presented for two smoothing scales: $\delta = \pi/2$ (middle row) and $\delta = \pi/4$ (bottom row).}
\end{figure}

\subsection{Ecliptic-Corrected \& CMB Dipole Removed Density Map}
Finally, we remove the CMB dipole signal from the ecliptic-corrected density map and reanalyze the data using the previously described methods. The results are shown in \autoref{fig:eclcor_nocmb_data}. Assuming we have successfully removed observational systematics, then the map (top left panel) should be statistically isotropic according to the standard cosmological model. However, after removing all known signals, residual large scale features are identified both in the density map and angular correlation function (top right panel). This signal, first observed in \cite{Secrest2021}, could be attributed to an anomalous dipole, { which suggests that the projected quasar distribution, when viewed as a two-dimensional field on the same celestial sphere as the CMB, does not share the same kinematic dipole as the CMB temperature map. Because the CatWISE sample effectively projects the three-dimensional quasar field onto the two-sphere through line-of-sight averaging, small-scale structure is smoothed and only the largest-scale (monopole and dipole) modes dominate. The statistical significance of this discrepancy is $\sim 5\sigma$, if we attribute the excess dipole reported by \cite{Secrest2021} purely to kinematic effects.} Our measurement of the angular correlation function of the data (top right panel) also presents a large scale dipole, whilst retaining a feature at $\theta \sim \pi/2$. 

\begin{figure*}[!htb]
\centering
\includegraphics[width=0.8\textwidth]{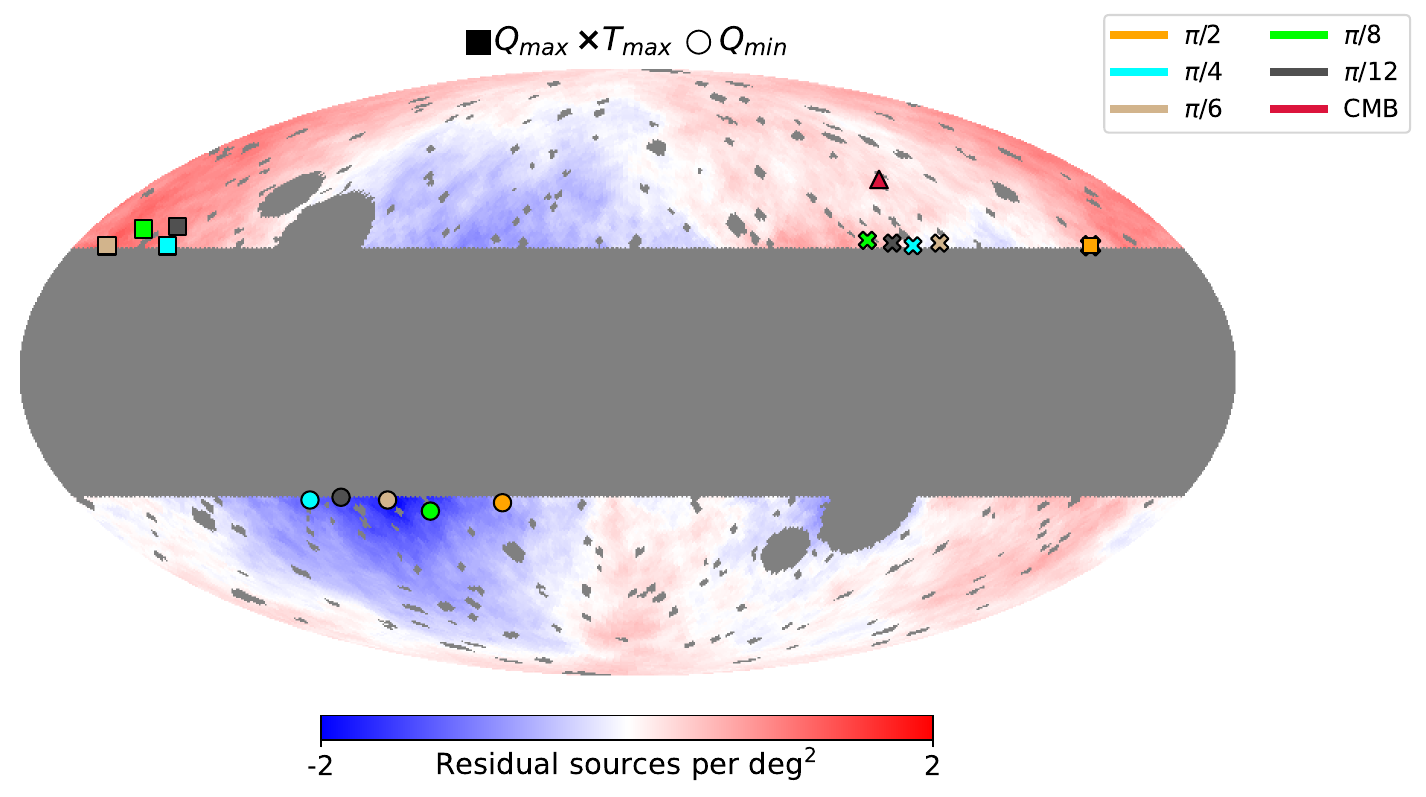}
\caption{\label{fig:direction}The directions of \Qmax, \Qmin and \Tmax on the sky, inferred from the ecliptic corrected, CMB dipole subtracted map smoothed over different scales. The squares/crosses/circles represent \Qmax/\Tmax/\Qmin, respectively. Results are presented for various smoothing scales from $\delta=\pi/2$ to $\pi/12$, using different colors. The quantities \Qmin and \Tmax are close to antipodal for most smoothing scales selected. The red triangle denotes the CMB dipole direction. The color map is the quasar data smoothed with a tophat of area 1 steradian, and is underplotted for reference.}
\end{figure*}

\begin{table}[ht!]
\centering

\begin{tabular}{|c|c|c|c|c|c|c|}
\hline
& $\pi/2$    & $\pi/4$    & $\pi/6$    & $\pi/8$ & $\pi/12$    & $\pi/16$    \\ \thickhline
\Qmax     & 0.00004   & 0.0003     & 0.0050    & 0.060  & 0.36  & 0.23         \\ \hline
\Qmin       & 0.00017     & 0.0002        & $1.2\times 10^{-6}$      & 0.0005 & 0.03     & 0.09  \\ \hline
\Tmax       & 0.00010     & 0.0054     & 0.0014    & 0.0057 & 0.03     & 0.25              \\ \thickhline
\end{tabular}
\caption{$p$-values for \Qmax, \Qmin and \Tmax of the field for various different smoothing scales $\pi/16 \leq \delta \leq \pi/2$. When smoothing on the largest scales, a dipole is clearly observed at high significance. As we decrease the smoothing, the minimum value of the field $Q_{\rm min}$ becomes the most significant feature in the map, down to $\delta = \pi/8$. As we continue to decrease $\delta$, the data becomes consistent with random fluctuations ($p$-values $>0.02$). }

\label{tab:q_columns}
\end{table}
\begin{figure*}[h]
\centering
    \begin{subfigure}[b]{\textwidth}
        \includegraphics[width=0.95\textwidth]{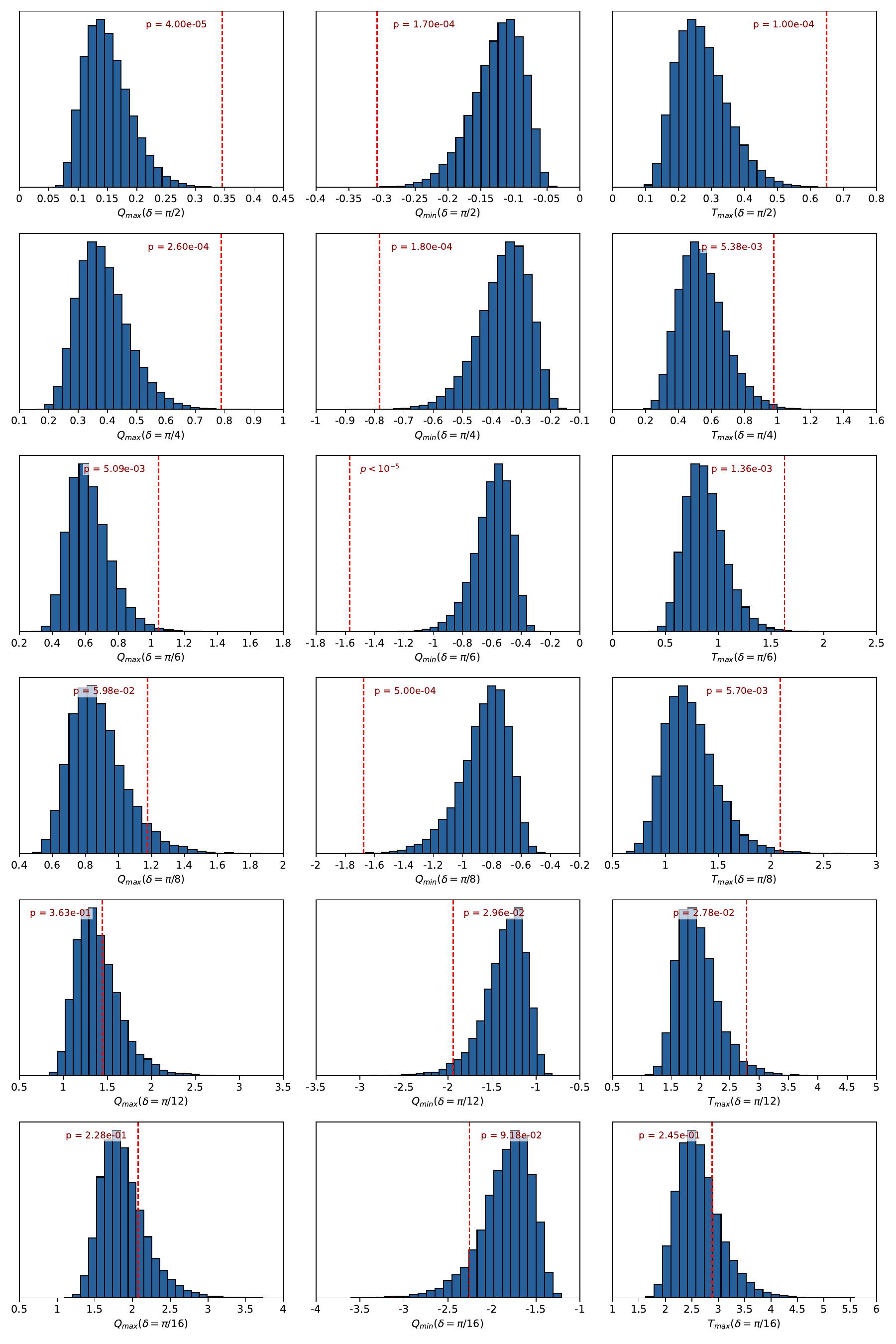}
    \end{subfigure}%
\caption{\label{fig:eclcor_nocmb_Qmax} The distributions of \Qmax, \Qmin and \Tmax from the mock catalog are shown alongside the corresponding values from the data {(vertical red dashed line)} for $\delta=\pi/2$ to $\pi/16$.}
\label{subfig:all_smooth}
\end{figure*}
\clearpage
Following the application of the ecliptic correction and the removal of the CMB dipole signal, the statistical significance of \Qmax, \Qmin, and \Tmax decrease markedly in comparison to the full density map. This is expected; we have removed the two largest sources of anisotropy. At both smoothing widths examined, the significance of \Qmax and \Tmax are practically the same, suggesting that the residual signal contains no significant quadrupolar contribution. These findings are presented in the middle and lower panels of \autoref{fig:eclcor_nocmb_data}.

We studied the map and associated summary statistics as a function of the smoothing scale $\delta$. As we decrease $\delta$, small scale modes will become increasingly dominant. If the anomalous signal in the data is due to large scale modes, then we can expect the significance of \Qmax, \Qmin and \Tmax to decrease with $\delta$. With this in mind, we apply smoothing at various scales down to $\delta = \pi/16$ and compute \Qmax, \Qmin and \Tmax summary statistics. The distributions of \Qmax, \Qmin and \Tmax for different smoothing scales are presented in \autoref{subfig:all_smooth} and the corresponding p-values of the data values relative to the random realisations are summarized in \autoref{tab:q_columns}. We note that the probability distributions are skewed, non-Gaussian and therefore we do not quote our results in terms of standard deviations. 

The statistical significance of \Qmax, \Qmin and \Tmax extracted from the data decreases with the decreasing smoothing scale, indicating that only the largest scales are anomalous. We find that \Qmin presents the most significant feature in the data, and its significance peaks at smoothing scale $\delta = \pi/6$  { with p-value \textit{$1.2\times 10^{-6}$}. This is similar in magnitude to the dipole significance reported in \cite{Secrest2021} }. We can see in the density map (cf top left panel of \autoref{fig:eclcor_nocmb_data}) that the under-density in the south sky is relatively localized compared to the overdense regions in the north, which are more diffuse. It is $Q_{\rm min}$ that drives the significance of $T_{\rm max}$ on scales $\delta \sim \pi/6$; the fact that the significance of \Tmax is lower than \Qmin indicates that a dipole is not the only feature in the data at these scales. As we decrease $\delta$ to values $< \pi/8$, the significance of all features in the data becomes negligible. This is expected, the small scale features in the map should be consistent with random fluctuations\footnote{There is clustering information in the quasar distribution \cite{Tiwari:2022hnf}, but we do not expect a significant correlation for the pixel sizes and smoothing scales selected in this work.}.  

\begin{figure}[t]
\begin{minipage}{.5\linewidth}
\centering
\includegraphics[width=.98\linewidth,valign=c]{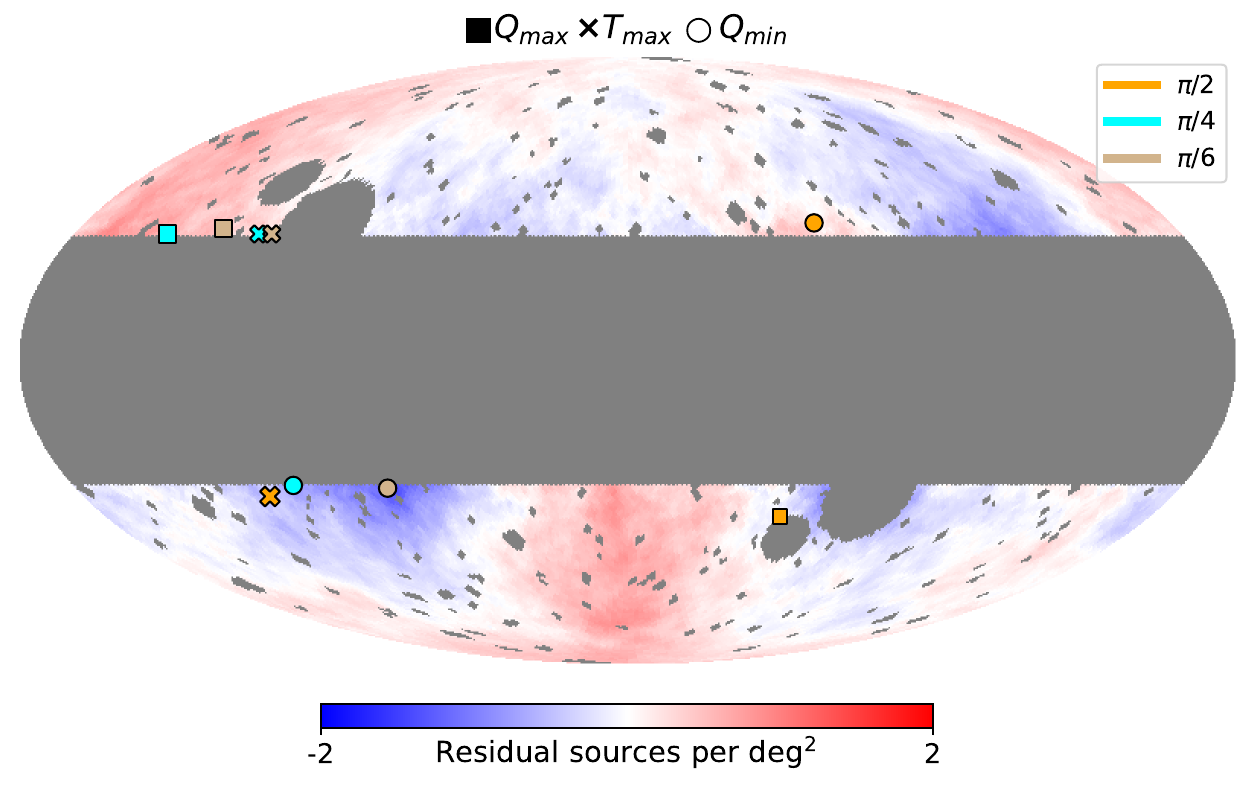}
\end{minipage}
\begin{minipage}{.5\linewidth}
\centering
        \includegraphics[width=0.92\textwidth]{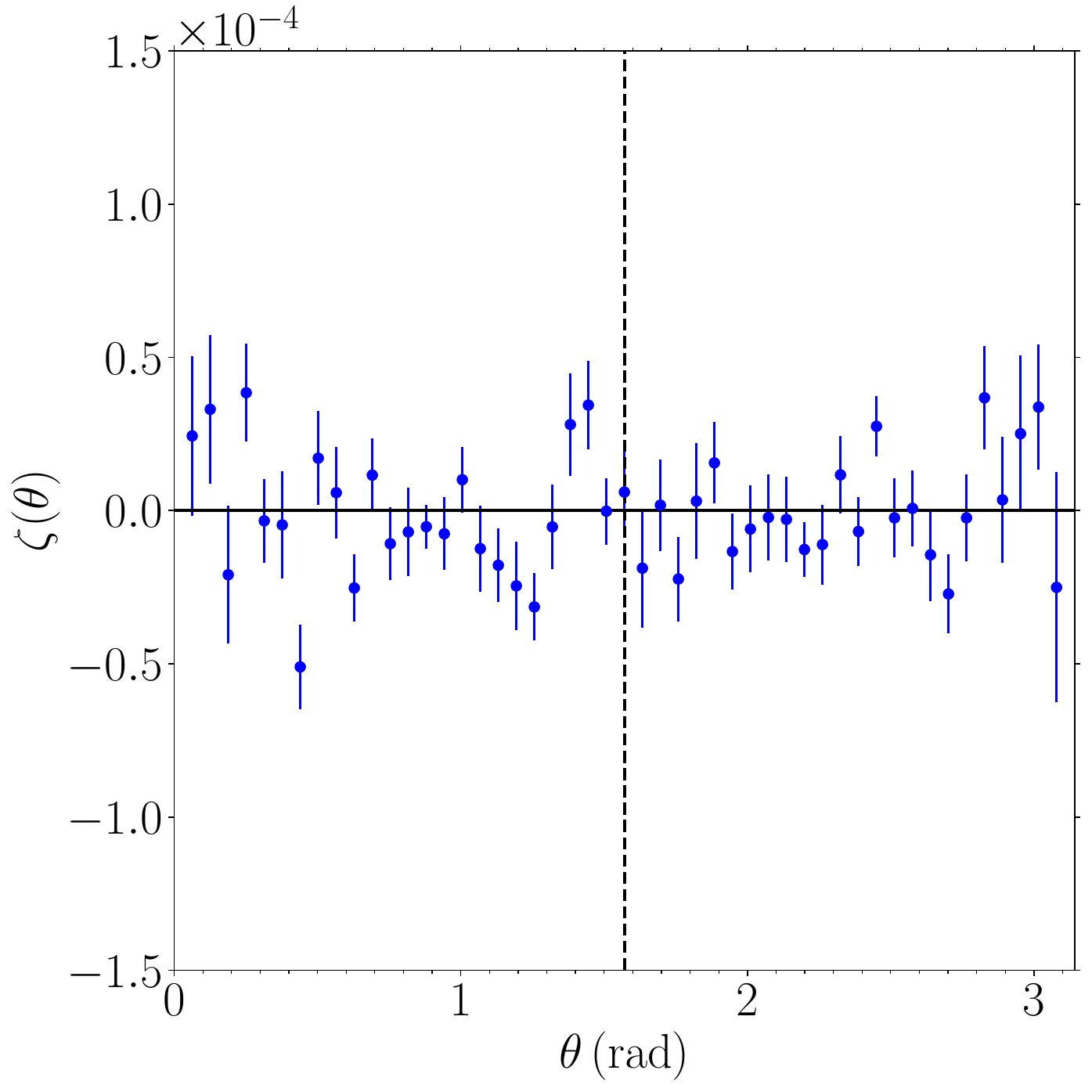}
\end{minipage} 
\\
        \includegraphics[width=0.92\textwidth]{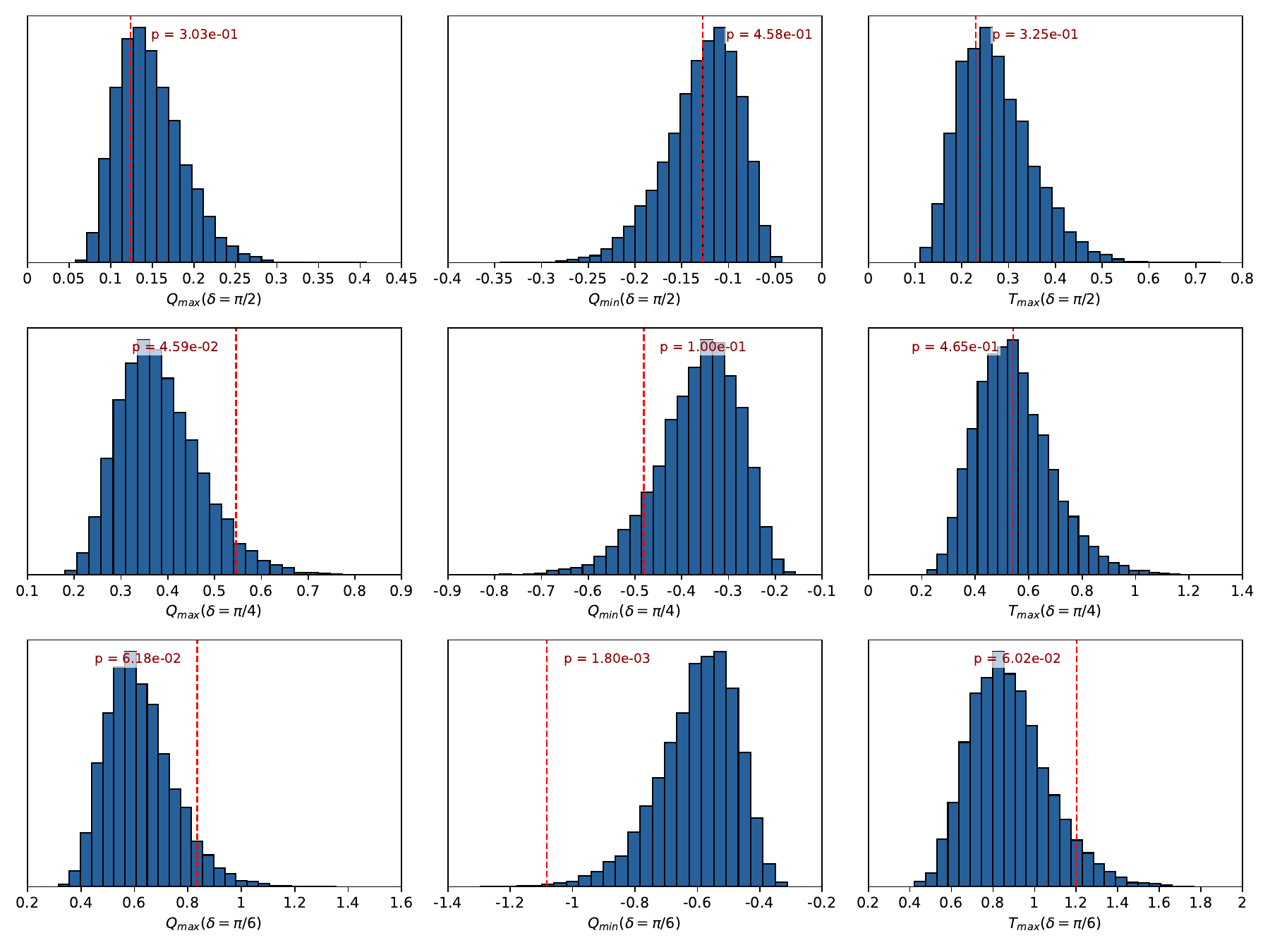}
\caption{\label{fig:eclcor_anom}Results from the real-space analysis of the ecliptic-corrected and anomalous dipole removed density map. 
\textbf{Top Left:} The smoothed density map of the quasar distribution from the CatWISE catalog. The gray regions represent masked areas in the map. 
\textbf{Top Right:} The angular correlation function, $\zeta(\theta)$, where $\theta$ is in radians. 
\textbf{Bottom Rows:} The distributions of \Qmax, \Qmin and \Tmax from $10^{5}$ isotropic realisations are shown (blue histograms) alongside the corresponding values of the same statistics measured from the data (vertical red dashed lines). Results are presented for three smoothing scales: $\delta = \pi/2$ (top) and $\delta = \pi/4$ (middle) and $\delta = \pi/6$ (bottom).}
\end{figure}

\subsection{Ecliptic-Corrected \& Anomalous Dipole Removed Density Map}
\label{sec:eclcor_anom}

The final case that we consider is the quasar map after it has been ecliptic corrected and the anomalous dipole inferred in \cite{Secrest2021} removed; ${\cal D} = 0.01554$, $(b,\ell) = (28.8^{\circ},238.2^{\circ})$. There is currently no theoretically compelling explanation for the existence of such a dipole; it is an empirical fit to the data after the quadrupole has been corrected. In this section, we determine the extent to which a pure dipole can remove the large scale anisotropic features from the ecliptic corrected map. The results are presented in Figure \ref{fig:eclcor_anom}.

In the top left panel, we still observe some large scale features in the map, including the southern under-dense region. This under-dense region is now surrounded by overdense patches, and the northern sky in the antipodal direction to $Q_{\rm min}$ is now under-dense. Both effects are a result of the dipole being an imperfect subtraction of the large scale features. We have shown that the significance of $Q_{\rm min}$ peaks when we smooth at $\delta = \pi/6$, whereas a dipole subtraction will modify the density field on scales $\sim \pi/2$. This is the reason why regions in the vicinity of $Q_{\rm min}$ now exhibit more pronounced overdensities. We have also found that $Q_{\rm max}$ is not typically antipodal to $Q_{\rm min}$, and so the dipole subtraction over-compensates in the north sky and creates a new large-scale under-density. However, in spite of these issues we note that the removal of the anomalous dipole has isotropized the map to a much higher degree than the CMB dipole subtraction in the previous section; the data are now mostly consistent with the random isotropic realisations as shown by the histograms in the lower panels. The only residual anisotropy of significance is $Q_{\rm min}$ when the map is smoothed at $\pi/6$ (cf. bottom middle panel) with corresponding p-value $p =0.0018$. This is the same under-density as found in the previous subsection. It remains present in the map, but with much reduced significance, because part of the signal has been attributed to the dipole and removed. 

{ Although reduced after the anomalous-dipole subtraction, this feature remains statistically noteworthy ($p=0.0018$), corresponding to roughly $3\sigma$. Its persistence suggests a genuine under-density or residual systematic localized in the vicinity $(b,\ell) = (-31^{\circ}, 78^{\circ})$, or likely at lower galactic latitude. Future spectroscopic surveys such as DESI will be able to determine whether this represents an actual structure or an observational artefact specific to the quasar data.}

\section{Discussion} 
\label{sec:discuss}

In this study, we generated smoothed density fields from CatWISE data and extracted the summary statistics \Qmax, \Qmin, and \Tmax. By comparing these values to those obtained from isotropic mock realizations, we assessed the statistical significance of anisotropy in the data across different smoothing scales. After accounting for known sources of anisotropy, we identified a residual anisotropic signal whose significance decreased as the smoothing width was reduced.

The quantities \Qmax and \Qmin are one-point functions, while \Tmax is a two-point function restricted to antipodal points on the sphere. Since our summary statistics differ from those used in previous studies, we do not expect to obtain the same statistical significance as dedicated dipole searches. In particular, \Qmin and \Qmax are local quantities that are not uniquely sensitive to a dipole. Conversely, \Tmax is sensitive to odd-multipole modes.  
Our results indicate the presence of a large-scale under-dense feature in the southern sky and more dispersed overdense patches in the north. Together, these features reveals the presence of a significant dipole (\Qmin and \Qmax are antipodal and both align with \Tmax) when the density field is smoothed on large scales (\(\delta \simeq \pi/2\)). However, when smoothed at $\delta \simeq \pi/2$, smaller scale modes will be generically suppressed relative to the dipole, and \Qmin, \Qmax are more likely to be antipodally located. On smaller scales, the under-dense patch remains significant down to \(\delta \sim \pi/8\) but \Qmin and \Qmax are no longer antipodal. As the smoothing width decreases further to \(\delta \lesssim \pi/8\), the data becomes fully consistent with isotropic, random realizations. Additionally, \Tmax is located in the region antipodal to \Qmin but not \Qmax, suggesting that \Tmax is predominantly driven by a minimum rather than a maximum-minimum pair in the smoothed data.  
Notably, we find a high significance for \Qmin at a smoothing width of \(\delta = \pi/6\) (p-value = \( 1.2\times 10^{-6} \)), pointing to the presence of a localized under-density in the data of characteristic size \(\sim \pi/6\) in the direction $(b,\ell) = (-31\degree,78\degree)$, with antipodal point $\sim (31\degree,258\degree)$. For comparison, if we smooth on large scales $\delta = \pi/2$, we find that \Qmin is antipodal to \Qmax with the latter located at $(b,\ell) = (30\degree,210\degree)$, in reasonable agreement with the result of \cite{Secrest2021}.

Finally, we measured the same statistics after removing the anomalous dipole inferred in \cite{Secrest2021} from the ecliptic-corrected quasar map. The removal of this dipole isotropizes the data to a higher degree than the CMB dipole subtraction, but some residual large scale features remain. This is unavoidable; we have found that maximally anisotropic features are present when we smooth on smaller scales $\delta \sim \pi/6$, which implies that any dipole subtraction must be imperfect in resolving this anisotropy. Even so, the anomalous dipole correction reduces the significance of any additional anisotropic features to p-values $p = 0.0018$. 

{ Our analysis points toward a local under-density patch rather than a strong dipole as reported by Secrest et al. The significance of this under-density, while evident across all smoothing scales, reaches its maximum at $\delta=\pi/6$, with a p-value of $1.2\times10^{-6}$. This value is comparable in magnitude to the p-value of the anomalous dipole reported in \cite{Secrest2021}, $p=5\times10^{-7}$.
Even when the anomalous dipole identified in \cite{Secrest2021} is accounted for and subtracted, our analysis continues to reveal an under-density patch with moderate significance. This suggests either (i) the presence of a local under-density region near the masked area in the southern hemisphere of the map, or (ii) a combination of an anomalous dipole accompanied by a local under-dense region.
While our study highlights the existence of this under-density feature, it does not rule out the anomalous dipole previously reported by Secrest et al.

}

{ The present CatWISE data lack redshift estimates, preventing a direct assessment of the depth of this under-density. Given its angular size, a cosmological-scale void interpretation appears unlikely, although it could be a low redshift artifact that presents in the full projected dataset. The method adopted in this work cannot locate features in masked regions, meaning the galactic latitude $b = -31^{\circ}$ represents a limiting value and the true value is likely to lie closer to the galactic plane. This could indicate a couple of possibilities. First,  there could be some residual stellar contamination in the quasar data. Despite the mid-infrared colour cut $W1 - W2 \geq 0.8$ specifically made to generate a clean and complete $(>95\%)$ extra-galactic sample, we expect some contamination, especially close to the galactic plane. Second, the feature lies in a region close to the local anisotropy found in the previous generation of type Ia Supernova catalogs \citep{Appleby:2014kea}. Although the significance of anisotropy in the supernova data was low, the alignment in direction with this work warrants further investigation.  Regardless, in the absence of redshift information we cannot provide any conclusive statements as to the origin of the feature here, but forthcoming spectroscopic data will allow this region of the sky to be re-examined.} 

The method adopted in this work can only find extreme field values in the `direction of unmasked pixels'; if the data contains a dipole that peaks within the mask then our approach will yield a biased estimate of the true direction. Different methods, such as directly fitting a dipole, can infer a preferred direction within masked regions\footnote{Direct fitting methods may also be prone to biases due to the presence of a mask and also due to prior assumptions made about the nature of the signal.}. However, our intention is not to specifically search for a dipole or even to find the preferred direction of any signal. Rather, our methodology is an attempt to search for anomalies in a model independent manner, by evaluating the significance of extreme values of the density field after smoothing at various scales. The method that we have adopted takes extreme values of the smoothed quasar density field on the sphere, and compares them to the same quantities extracted from isotropic random resampling of the data. The methodology can be used to test for dipoles and other large scale modes, by comparing the direction and magnitude of \Qmax, \Qmin and \Tmax as a function of smoothing scale. 

The quasar data contains large scale modes that are statistically rare compared to random, isotropic re-sampling of the same data. The exact nature and origin of these modes, and whether they are sourced by unknown systematics or physical effects beyond the standard cosmological model, remains an open question and will require future study. Of particular interest is the determination of the redshift dependence of the signal, and also its sensitivity to other properties of the data -- brightness, color etc. Confirming the signal in other data, such as done in \cite{Secrest:2022uvx}, is also an important future goal. Finally, studying the properties of other large scale structure catalogs could shed more light on the structures found in this work.

\acknowledgments
The authors would like to thank Roya Mohayaee, Subir Sarkar and Nathan Secrest for insightful discussions. AA and SA are supported by an appointment to the JRG Program at the APCTP through the Science and Technology Promotion Fund and Lottery Fund of the Korean Government, and were also supported by the Korean Local Governments in Gyeongsangbuk-do Province and Pohang City. SA also acknowledges support from the NRF of Korea (Grant No. NRF-2022R1F1A1061590) funded by the Korean Government (MSIT). A.S. would like to acknowledge the support by National Research Foundation of Korea 2021M3F7A1082056, and the support of the Korea Institute for Advanced Study (KIAS) grant funded by the government of Korea.

\appendix

\section{Appendix} 

\subsection{Removing a dipole signal from a radio source map}\label{dipole_removal}

Ellis and Baldwin derived the dipole anisotropy in the number count of radio sources that are isotropically distributed when observed from a reference frame moving with velocity $v$ \cite{Ellis1984}. 
Consider a population of radio sources with power-law spectra, given by $S \propto \nu^{-\alpha}$, 
and an integral source count distribution per unit solid angle expressed as  
$
    \frac{dN}{d\Omega}(>S) \propto S^{-x},
$
above a limiting flux density $S$. When viewed from a moving reference frame with velocity $v$, this distribution exhibits a dipole anisotropy with an amplitude given by  
\begin{equation}
    \mathcal{D} = [2 + x(1+\alpha)]\frac{v}{c}.
\end{equation}

\noindent Since our local frame moves with a velocity of $v = 369$ km/s relative to the cosmic microwave background (CMB) rest frame, a dipole anisotropy of amplitude $\mathcal{D}_{\text{CMB}} = 0.007$ is expected in the number count of radio sources, aligned with the direction of motion. To quantify any excess dipole signal, this CMB-induced dipole must be subtracted.  

Given a full-sky distribution, a dipole anisotropy of amplitude $\mathcal{D}$ along a direction $\hat{d}$ can be introduced or removed using the transformation  
\begin{equation}
    S(\hat{n}) = S_0 \left( 1 \pm \mathcal{D} \,\hat{n} \cdot \hat{d} \right),
\end{equation}
where $\hat{n}$ denotes the unit vector corresponding to a given sky position.

\clearpage

\bibliography{main.bib}

\end{document}